\documentclass{emulateapj}
\usepackage{amsmath}
\usepackage{color}
\usepackage{ulem}

\shorttitle{The VMC Survey. XXIX. Hierarchical Star Formation in the SMC}
\shortauthors{Sun et al.}
\slugcomment{Submitted for Publication in The Astrophysical Journal}

\begin{document}

\title{The VMC Survey. XXIX. Turbulence-Controlled Hierarchical Star Formation in the Small Magellanic Cloud}

\author{Ning-Chen~Sun$^{1}$, 
Richard~de~Grijs$^{1, 2, 3}$, 
Maria-Rosa~L.~Cioni$^4$,
Stefano~Rubele$^{5, 6}$,
Smitha~Subramanian$^7$,
Jacco~Th.~van~Loon$^8$,
Kenji~Bekki$^9$,
Cameron~P.~M.~Bell$^4$,
Valentin~D.~Ivanov$^{10, 11}$,
Marcella~Marconi$^{12}$,
Tatiana~Muraveva$^{13}$,
Joana~M.~Oliveira$^{10}$,
Vincenzo~Ripepi$^{12}$}
\affil{
$^1$ Kavli Institute for Astronomy \& Astrophysics and Department of Astronomy, Peking University, Yi He Yuan Lu 5, Hai Dian District, Beijing 100871, China; sunnc@foxmail.com, grijs@pku.edu.cn \\
$^2$ Department of Physics and Astronomy, Macquarie University, Balaclava Road, Sydney NSW 2109, Australia \\
$^3$International Space Science Institute -- Beijing, 1 Nanertiao, Hai Dian District, Beijing 100190, China \\
$^4$ Leibniz-Institut f\"ur Astrophysik Potsdam, An der Sternwarte 16, 14482 Potsdam, Germany \\
$^5$ INAF-Osservatorio Astronomico di Padova, vicolo dell'Osservatorio 5,  I-35122, Padova, Italy \\
$^6$ Dipartimento di Fisica e Astronomia, Universit\`a di Padova, vicolo dell'Osservatorio 2, I-35122, Padova, Italy \\
$^7$ Indian Institute of Astrophysics, Koramangala II Block, Bangalore-34, India \\
$^8$ School of Chemical \& Physical Sciences, Lennard-Jones Laboratories, Keele University, ST5 5BG, UK \\
$^9$ ICRAR, M468, The University of Western Australia 35 Stirling Highway, Crawley Western Australia, 6009, Australia \\
$^{10}$ ESO European Southern Observatory, Ave. Alonso de Cordova 3107, Casilla 19, Chile \\
$^{11}$ ESO Garching: ESO,  Karl-Schwarzschild-Str. 2, 85748 Garching bei M\"unchen, Germany  \\
$^{12}$ INAF-Osservatorio Astronomico di Capodimonte, Salita Moiariello 16, I-80131 Napoli, Italy \\
$^{13}$ INAF-Osservatorio Astronomico di Bologna, via Ranzani 1, I-40127, Bologna, Italy \\
}

\begin{abstract}

In this paper we report a clustering analysis of upper main-sequence stars in the Small Magellanic Cloud, using data from the VMC survey (the VISTA near-infrared $YJK_s$ survey of the Magellanic system). Young stellar structures are identified as surface overdensities on a range of significance levels. They are found to be organized in a hierarchical pattern, such that larger structures at lower significance levels contain smaller ones at higher significance levels. They have very irregular morphologies, with a perimeter--area dimension of 1.44~$\pm$~0.02 for their projected boundaries. They have a power-law mass--size relation, power-law size/mass distributions, and a lognormal surface density distribution. We derive a projected fractal dimension of 1.48~$\pm$~0.03 from the mass--size relation, or of 1.4~$\pm$~0.1 from the size distribution, reflecting significant lumpiness of the young stellar structures. These properties are remarkably similar to those of a turbulent interstellar medium (ISM), supporting a scenario of hierarchical star formation regulated by supersonic turbulence.

\end{abstract}

\keywords{galaxies: star clusters -- infrared: stars -- Magellanic Clouds -- stars: formation}

\section{Introduction}

Star formation leads to hierarchical young stellar structures ($\lesssim$~10$^8$~yr), including star clusters, associations, and complexes of increasing size and decreasing density \citep{Ivanov1992, Efremov1995, Efremov1998, Bastian2006, Bastian2009b}. A complex may be subclustered into star clusters and associations \citep{Eigenson1988, Piskunov2006, de la Fuente Marcos2008, Elmegreen2011}, and the associations may in turn contain a number of subgroups and star clusters \citep[e.g. the Orion~OB1 association;][]{Genzel1989, Kuhn2014, Da Rio2014}. The young stellar structures are a natural outcome of star formation, which itself is a hierarchical process and appears correlated in both space and time. This is supported by the age difference--separation relation of star clusters \citep{Efremov1998, Grasha2017a}, young star clusters forming in pairs or groups \citep{Bhatia1988, Hatzidimitriou1990}, as well as the positional correlations of newly-formed stars \citep{Gomez1993, Larson1995, Gieles2008, Bastian2009, Gouliermis2015} and star clusters \citep{Zhang2001, Scheepmaker2009, Grasha2017b}. Interstellar turbulence may play a key role in regulating hierarchical star formation across a wide scale range \citep{Efremov1998}, possibly aided by fragmentation with agglomeration and self gravity \citep{Carlberg1990, deVega1996}. The young stellar structures undergo rapid dynamical evolution after their formation. Most of the newly formed star clusters are short-lived, and only a small fraction are able to survive the early disruptive processes \citep{Lada2003, PortegiesZwart2010}. On larger scales, stellar associations and complexes are not gravitationally bound, and they dissolve into the general galactic field within no more than several hundred million years \citep{Efremov1995, Gieles2008, Bastian2009, Gouliermis2015}. It is therefore important to study their properties, formation and evolution, which will provide insights into a galaxy's star formation and dynamical processes on multiple scales.

Located at a distance of $\sim$62~kpc \citep[e.g.][]{Cioni2000a, deGrijs2015, Ripepi2016}, the Small Magellanic Cloud (SMC) serves as an important astrophysical laboratory. It is undergoing dynamical interactions with the Large Magellanic Cloud \citep[LMC;][]{Irwin1990, Bekki2007, Belokurov2017, Ripepi2017, Subramanian2017} as well as with the more distant Milky Way \citep{Besla2007, Diaz2011, DOnghia2016}. It is abundant in both current and past star-forming activity \citep{Oliveira2009, Rubele2015}, and its stellar populations can be spatially resolved even with ground-based telescopes. The star cluster population of the SMC has been extensively investigated \citep[e.g.][]{Pietrzynski1998, deGrijs2008, Gieles2008, Piatti2016}. In contrast, its stellar associations have received far less attention [see Table~4 of \citet{Bica2008} for a catalog of SMC associations]. Note that the SMC is too small to contain many stellar complexes. The young stellar structures in the SMC are thus poorly understood especially in terms of their hierarchical nature.

The early work of \citet{Hodge1985} reported 70 OB associations in the SMC. With a mean diameter of 77~pc, their SMC associations have a similar mean size as those in the LMC, but are smaller than those in the Milky Way, M31, and M33. As pointed out by the author, however, this may not be a true size difference and might result from selection effects -- the different abilities to recognize associations in galaxies at different distances, and the different criteria, often subtle and subjective, to separate the stellar distribution into associations. \citet{Battinelli1991} proposed an objective method, which they used to identify associations in the SMC. They reported a mean diameter of 90~pc for the associations, which is comparable to that of \citet{Hodge1985}. However, as \citet{Bastian2007} pointed out, this characteristic size may also be a selection bias: the mean size would change when adopting different values for the breaking scale and minimum number of stars in the identification algorithm. Actually, the stellar associations may not have any characteristic sizes.

\citet{Gieles2008} investigated the spatial distributions of stars of various ages in the SMC. They found that stars are born with significant substructures in their spatial distribution with a fractal dimension of 2.4. Older stars become successively less subclustered, and the stellar distribution becomes smooth on a timescale of $\sim$75~Myr. Unfortunately, they did not investigate the properties of the young stellar structures in greater detail. \citet{Bonatto2010} studied the size distributions of star clusters and ``non-clusters" (nebular complexes and stellar associations) in the Magellanic Clouds. They found that the size distributions follow power laws, which is in agreement with a scenario of hierarchical star formation. However, they relied on the catalog of extended sources of \citet{Bica2008}, which is in turn based on the extensive previous catalogs from different authors. Thus, it is not easy to assess the effects of selection biases and incompleteness.

It is therefore not trivial to identify young stellar structures~\footnote{In the following we shall not distinguish star clusters, associations, and complexes, and we choose to refer to all of them as young stellar structures.} while avoiding subjectivity and selection bias. One important reason comes from their hierarchical nature, in the sense that they have irregular morphologies and abundant substructures -- they themselves may be the substructures of larger ones as well \citep[see also the discussion by][]{Elmegreen1998}. This calls for systematic, objective and unbiased identification algorithms. A contour-based map analysis technique, proposed by \citet{Gouliermis2015, Gouliermis2017}, has proved to be effective to achieve this end. The idea is to identify young stellar structures as overdensities on a {\it series} of significance levels from the surface density map of young stars. This method provides an intuitive view of the young stellar structures and, in particular, their hierarchical nature. This technique, or similar ones based on the same principles, have been applied to a number of nearby galaxies \citep[e.g.][]{Gouliermis2010, Gouliermis2015, Gouliermis2017} and star-forming complexes \citep[e.g.][]{Sun2017a, Sun2017b}. Still, it has not yet been applied to the SMC as a whole.

Considering that the young stellar structures span multiple scales, the observations require both high spatial resolution and wide-field coverage. The VMC survey \citep{Cioni2011} provides a precious opportunity to meet this challenge. With a spatial resolution of $\lesssim$1$\arcsec$ (0.3~pc at the SMC's distance), it covers both Clouds' full classical extents at $B$~=~25~mag~arcsec$^{-2}$ \citep{Bothun1988}, the Bridge between the two galaxies, with two additional fields centered on their associated Stream. Working at near-infrared wavelengths, the VMC survey is less susceptible to the effects of reddening, which is usually important for the young stellar populations. In this paper, we apply the contour-based map analysis to the SMC, using upper main-sequence (UMS) stars from the VMC survey. Our goal is to understand the hierarchical properties of its young stellar structures, which will also demonstrate their formation mechanism(s). In Section~\ref{data.sec} we outline the VMC data used in this work, and Section~\ref{map.sec} describes how we build the stellar density map. Young stellar structures are identified in Section~\ref{id.sec}. We analyze their properties in Section~\ref{property.sec}, followed by a discussion in Section~\ref{discussion.sec}. We finally close this paper with a summary and conclusion, leaving the issue of their evolution for a future paper.

\section{Selected VMC Data}
\label{data.sec}

Data used in this work come from the VMC survey, which is carried out with the VISTA telescope \citep{Sutherland2015}. We refer to \citet{Cioni2011} for a comprehensive description of the VMC survey. The original pawprints\footnote{Because there are gaps between the detectors, VISTA observes a contiguous area of sky using a sequence of six offset observations. Each of the six observations is called a {\it pawprint}, while the combined image, covering a contiguous area of $\sim$1.5~deg$^2$, is referred to as a {\it tile}.} are processed by the VISTA Data Flow System \citep[VDFS;][]{Irwin2004}, and PSF photometry is performed based on stacked tiles combining all good-quality epochs \citep{Rubele2012, Rubele2015}. We use PSF photometry in this work, since it suffers less from source crowding than aperture photometry \citep{Tatton2013}. The data are retrieved from the VISTA Science Archive~\footnote{\url{http://horus.roe.ac.uk/vsa}} \citep[VSA;][]{Cross2012}.

\begin{figure}
\centering
\includegraphics[scale=0.44,angle=0]{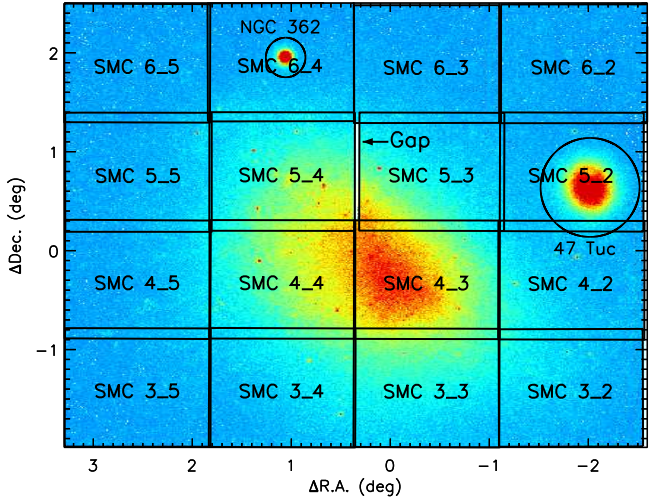}
\caption{VMC tiles used in this work. The background is a density map of VMC sources with $K_s$~$<$~20~mag, with red colors for high densities and blue colors for low densities. The two circles indicate the globular-cluster regions which are excluded from our analysis.}
\label{map.fig}
\end{figure}

There are 27 SMC tiles in the VMC survey. In this paper, we only use the central 4~$\times$~4 tiles covering the main body of the SMC and its close vicinity. The other tiles are in the more outer regions and have very low stellar surface densities; thus they are not analyzed in this work. The names and extents of the 16 tiles for analysis are shown in Fig.~\ref{map.fig} overlaid on a stellar density map. Tiles SMC~3\_3, 3\_5, 4\_3, 5\_2, and 5\_4 are part of Data Release 4 while the other tiles are currently still proprietary to the VMC team (they will be included in Data Release 5 expected in 2018). The stellar density map of Fig.~\ref{map.fig} is obtained from simple star counts of VMC sources with $K_s$~$<$~20~mag in 0.01$^\circ$~$\times$~0.01$^\circ$ cells. This magnitude cut includes the UMS, red giant branch (RGB), and the red clump (RC) of the SMC. The SMC's main body is clearly revealed in the central region of the map. Tiles SMC~3\_5, 4\_5, and 5\_5 cover the inner part of the SMC's Wing, which is located to the east of the main body. It is obvious that the 16 tiles contain the bulk of the SMC's stars.

Adjacent tiles partly overlap with each other; as a result, there are duplicate sources in the combined source catalog. If a source is located in the overlap region and has detections in two or more tiles, we keep only the detection from the tile whose center is closest to the source's position. In this way, the duplicate sources are removed and all sources are unique in the combined catalog. According to the tiling algorithm of VISTA, strips along the upper and lower outer edges of each tile are not covered to the same depth as the bulk of the tile. This is because only one pawprint contributes there, while the bulk of the tile is covered by 2--6 pawprints. Ideally, one needs to coadd the objects in the overlap regions between two tiles to obtain uniform depth. However, this effect does not make any significant difference in the context of this work, since the sources under investigation are very bright compared with the survey's magnitude limits (see Section~\ref{ums.sec}).

Two overdensities are located in tiles SMC~5\_2 and 6\_4. They correspond to the Galactic globular clusters, 47~Tucanae (47~Tuc) and NGC~362. In the following analysis, we exclude all sources within 0.5$^\circ$ and 0.2$^\circ$ from their centers\footnote{47~Tuc is centered at R.A.(J2000)~=~00$^{\rm h}$24$^{\rm m}$04$^{\rm s}$.8, Dec.(J2000)~=~$-$72$^\circ$04$\arcmin$48$\arcsec$, while NGC~362 is centered at R.A.(J2000)~=~01$^{\rm h}$03$^{\rm m}$14.3$^{\rm s}$.8, Dec.(J2000)~=~$-$70$^\circ$50$\arcmin$56$\arcsec$.}, respectively, which can effectively eliminate the influence of the two globular clusters. The stellar density map also shows a vertical strip devoid of stars at $\Delta$R.A.~$\sim$~0.3$^\circ$. This arises from a gap between tiles SMC~5\_3 and 5\_4. Since SMC~5\_3 is slighted offset to the west, this gap is not covered by the observations. A complementary VISTA observation is ongoing to fill this gap. In this work, we will use the Magellanic Cloud Photometric Survey \citep[MCPS;][]{Zaritsky2000, Zaritsky2002} to fill this gap, as detailed in Section~\ref{mcps.sec}.

\section{KDE Map of the UMS Stars}
\label{map.sec}

\subsection{Selection of UMS Stars}
\label{ums.sec}

\begin{figure*}[htb!]
\centering
\begin{tabular}{cc}
\includegraphics[scale=0.6,angle=0]{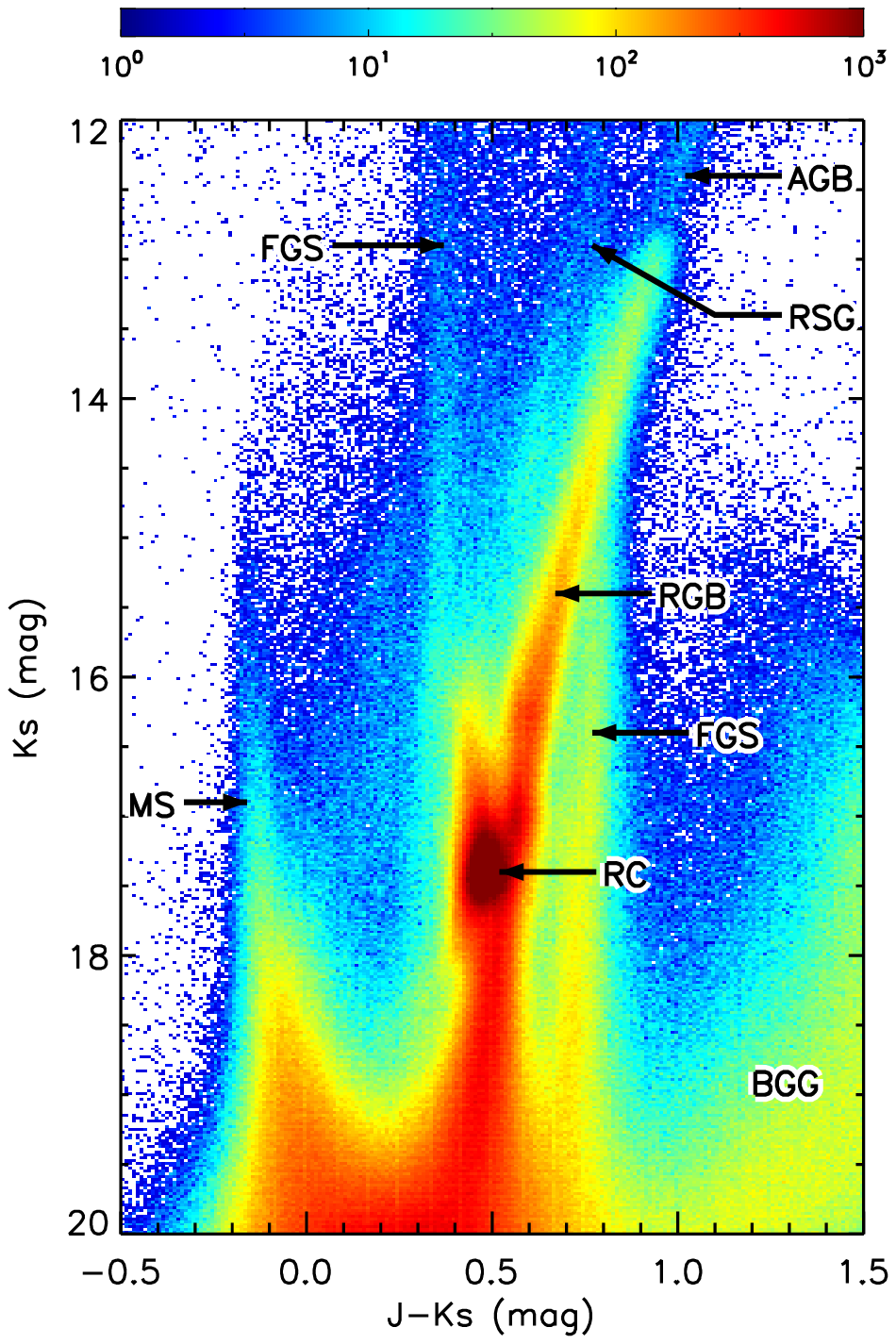}
\includegraphics[scale=0.6,angle=0]{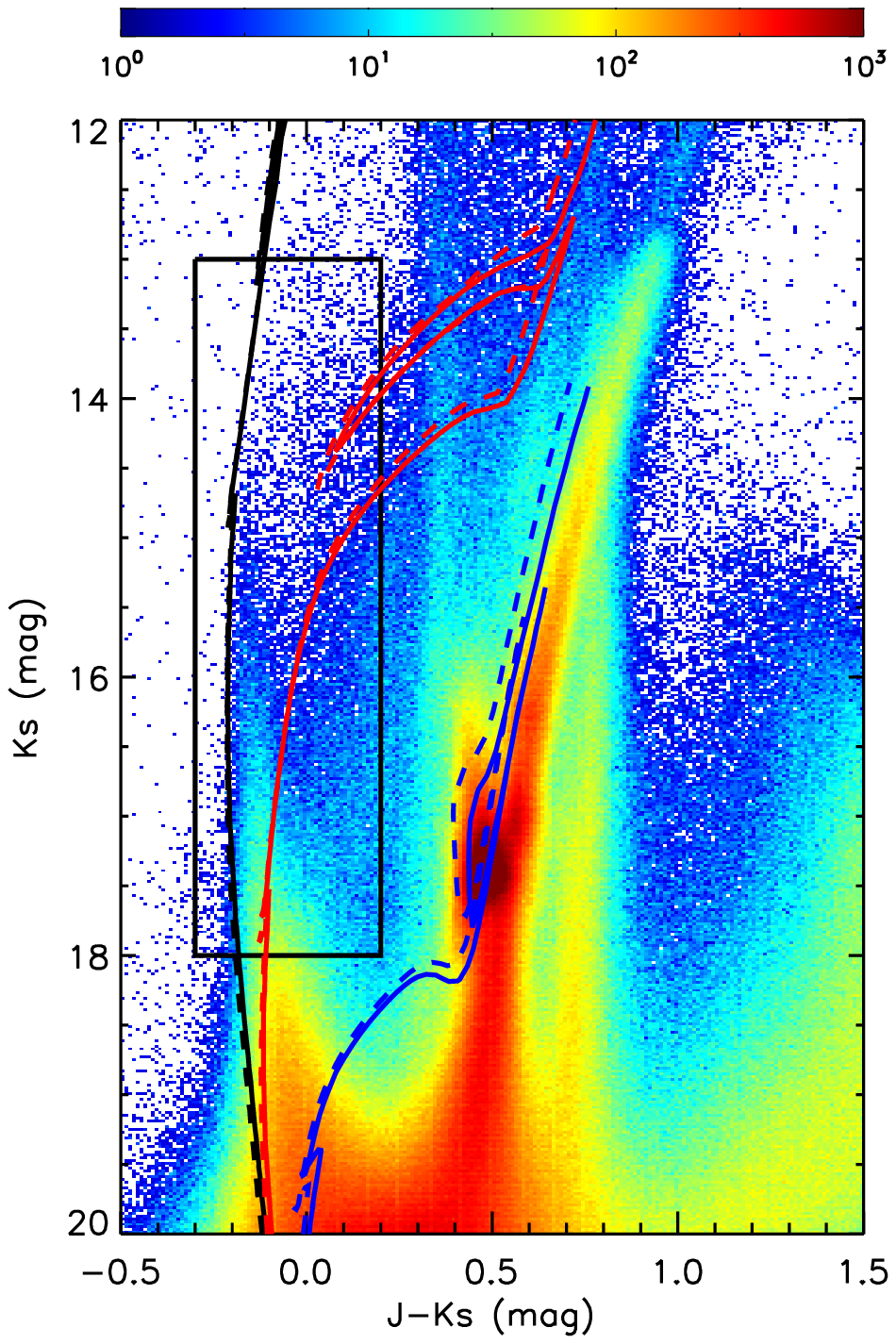}
\end{tabular}
\caption{($J - K_s$, $K_s$) color--magnitude Hess diagram of sources in the 16 selected SMC tiles. The color scales show the number of stars in each color--magnitude bin, with a bin size of 0.01~mag in color and 0.02~mag in magnitude. In the left-hand panel, a number of apparent features are labeled (see text). In the right-hand panel, PARSEC (version~1.2S) isochrones are overplotted, shifted by a distance modulus of ($m$~$-$~$M$)$_0$~=~18.96~mag and a foreground Galactic extinction of $A_V$~=~0.12~mag. They correspond to ages of log($\tau$/yr)~=~7.0 (black), 8.0 (red), and 9.0~dex (blue) with 20\% or 10\% solar metallicity (solid or dashed lines, respectively). The black box shows the criteria adopted for selecting UMS stars. The color scales in both panels are identical; the labels, isochrones, and selection box are shown separately simply for clarity.}
\label{cmd.fig}
\end{figure*}

Figure~\ref{cmd.fig} shows the ($J - K_s$, $K_s$) color--magnitude diagram (CMD) of all sources in the 16 selected SMC tiles. The main sequence (MS), RGB, and RC are all clearly visible. Brighter than the tip of the RGB is the asymptotic giant branch (AGB), and the red supergiant branch (RSG) is located slightly bluer than the RGB and AGB. Redward of $J - K_s$~=~1.0~mag are background galaxies (BGGs), and the vertical spurs at $J - K_s$~=~0.75~mag and 0.35~mag are foreground stars (FGSs) in the Milky Way.

In the right-hand panel of Fig.~\ref{cmd.fig}, we also show the PARSEC \citep[version~1.2S;][]{Bressan2012} isochrones of ages log($\tau$/yr)~=~7.0, 8.0, and 9.0. The solid and dashed isochrones correspond to 20\% and 10\% of solar metallicity, respectively\footnote{The PARSEC isochrones have adopted a solar metallicity of $Z_\odot$~=~0.0152, and are computed following a $Y$~=~0.2485~+~1.78$Z$ relation for the helium abundance. Here $Y$ and $Z$ are the mass fractions of helium and metal elements.}. The former metallicity is derived from H~{\scriptsize II} regions and young stars \citep{Russell1992}, while the latter is typical of the RGB population in the SMC \citep{Dobbie2014}.  Offsets of 0.026~mag in $J$ and 0.003~mag in $K_s$ have been subtracted from the isochrones to convert the model magnitudes from the Vega system to the VISTA system \citep{Rubele2012, Rubele2015}. The isochrones are shifted by an SMC distance modulus of ($m - M$)~=~18.96~mag \citep{deGrijs2015} and corrected for a foreground Galactic extinction of $A_V$~=~0.12~mag \citep{Schlegel1998}. We adopt the extinction coefficients $A_J$/$A_V$~=~0.283 and $A_{K_s}$/$A_V$~=~0.114, which are computed from the \citet{Cardelli1989} extinction curve with $R_V$~=~3.1 \citep{Girardi2008}. The extinction coefficients in $J$ and $K_s$ depend only very weakly on the stellar spectral types or the adopted extinction laws \citep[see e.g.][]{vanLoon2003}.

UMS stars are selected based on 13.0~$<$~$K_s$~$<$~18.0~mag and $-$0.3~$<$~$J - K_s$~$<$~0.2~mag, which is indicated by the black solid box in Fig.~\ref{cmd.fig}. The isochrones suggest that the selected stars are younger than 1~Gyr. Since the VMC survey has typical saturation limits of $J$~=~12.7~mag and $K_s$~=~11.4~mag \citep{Cioni2011}, the upper limit of $K_s$~=~13.0~mag ensures that the selected stars are not saturated in either band. This limit corresponds to a stellar mass of $\sim$19~M$_\odot$ for an age of 10~Myr and a 20\% solar metallicity. The color range is chosen with consideration of interstellar extinction. In addition to the foreground Galactic extinction, the stars also suffer from extinction inside the SMC. This component is, however, very difficult to model theoretically or derive observationally. First, extinction in star-forming regions often exhibits significant spatial variations \citep[e.g.][]{Lombardi2010}. Second, dust may lie in the foreground or background of a star, and only the foreground dust contributes to its extinction. Thus, an extinction map derived from a group of stars is not always directly applicable to another group of stars, if they have different distributions along the line of sight with respect to the dust. Moreover, extinction in the Magellanic Clouds exhibits a population dependence \citep{Zaritsky1999}. Extinction has a very small effect on the infrared magnitudes, but may shift some UMS stars redward out of the selection box. Thus, we have adopted a wide color range in the selection criteria to include the reddened stars as completely as possible. For instance, an extinction of $A_V$~=~1.8~mag is needed to shift a star redward from the log($\tau$/yr)~=~7.0~dex isochrone out of the selection box; such a high extinction is very rare in the SMC \citep[see e.g.][]{Haschke2011, Rubele2015}. The UMS sample contains 46,148 stars.

Because we have adopted a wide color range in the selection, the UMS sample consists of both UMS stars and slightly evolved stars that are moving redward off the MS or on the blue loops. However, their number is likely small, since a star does not spend much time on these rapid evolutionary phases. On the other hand, many of these stars may still have very young ages; if so, they can also help trace the young stellar structures.

\citet{Rubele2018} analyzed the star formation history (SFH) of the SMC based on a stellar population synthesis technique. Their mapped region is the same as that adopted in this work. They found that the SMC had a very low star-forming rate at log($\tau$/yr)~=~8.5--9.0 (lookback time), while more recently it has experienced increased star-forming activity. Based on their SFH, it is possible to assess the age distribution of stellar populations in the SMC (see their Section~5.4). We have done such an analysis for the UMS sample based on their results. We found that the sample has a median age close to log($\tau$/yr)~=~7.7 (50~Myr); approximately 72\%, 86\%, and 96\% of the selected stars are younger than log($\tau$/yr)~=~8.0, 8.2, and 8.4 (100, 160, and 250~Myr), respectively. Thus, while the comparison with isochrones (Fig.~\ref{cmd.fig}) gives an upper age limit of 1~Gyr, a detailed SFH analysis suggests a much younger age distribution for the UMS sample.

Here we also make an important distinction between the ages of the selected UMS stars and the ages of the young stellar structures. In this work, the young stellar structures will be identified as surface overdensities (Section~\ref{id.sec}). On the other hand, recent papers have shown that young stars are the most subclustered while old ones have very smooth spatial distributions \citep{Gieles2008, Bastian2009, Gouliermis2010, Sun2017b}. This is because the young stellar structures undergo rapid, disruptive dynamical evolution and/or because of the spatial overlap of different generations of star formation \citep{Elmegreen2018}. As a result, the surface overdensities are dominated by young rather than old stars.

\citet{Gieles2008} analyzed the age-dependent two-point correlation functions of the SMC's overall stellar distributions. They found that the stellar distribution becomes indistinguishable from a uniform distribution at an age of 75~Myr. Although this value cannot be taken as an upper age limit for a specific individual stellar structure, statistically speaking most stellar structures should be dispersed roughly on this timescale (otherwise, the stellar distribution would still be highly non-uniform). \citet{Bastian2009} derived a timescale of 175~Myr for the LMC, and \citet{Sun2017b} obtained a timescale of 100~Myr for the LMC's Bar complex~\footnote{Note the derived timescales could suffer from significant uncertainties due to large age spreads of the used stellar samples \citep[see the discussion in][]{Sun2017b}.}. On the other hand, most young stellar structures are unbound, except for some compact star clusters on small scales. They could disperse due to the stellar random motions, which are related to the turbulent motions in their natal star-forming clouds. The turbulent crossing time is typically several tens of millions of years for a cloud of 100~pc \citep[see e.g.][]{Efremov1998}. The structure dispersion timescale should be comparable to or a few times the turbulent crossing time, if stellar random motions dominate their dispersion process (this is a valid assumption especially for large-scale, loosely bound or unbound structures). In addition, the recent LMC--SMC encounter at 200--400~Myr ago must have had a strong impact on the SMC \citep{Maragoudaki2001, Bekki2007}; some of the stellar structures older than this age could have been more easily destroyed under the LMC's tidal force. Considering all this, it is reasonable to assume that most young stellar structures have an upper age limit of the order of 100~Myr.

One may still question whether the results can be affected by contamination of stars in the UMS sample which are older than the typical ages of young stellar structures. We have done a test with two additional UMS samples, each selected with a different fainter magnitude cut at $K_s$~=~17.5 or 18.5~mag. They have different stellar age distributions compared with the original sample. 77\% or 58\% of their stars are younger than 100~Myr, respectively, so they have a reduced or increased fraction of older stars. We then repeat the entire analysis as described below; the results do not change significantly and the derived quantities (perimeter--area dimensions, slopes of size/mass distributions, fractal dimension, etc.; see Section~\ref{property.sec}) are consistent within the uncertainties. This test again confirms that the young stellar structures are dominantly revealed by the young rather than the old stars. Thus, the conclusions reached below are robust and not affected by the contamination by old stars.

In the sample selection we have adopted a single distance modulus for the SMC. However, the SMC is known to be significantly elongated by more than 25--30~kpc along the line of sight \citep[e.g.][]{Subramanian2009, Ripepi2017}. As will be shown later, our identified young stellar structures lie primarily in the central main body of the SMC. The SMC's line-of-sight depth is much smaller in this limited area. On the other hand, \citet{Ripepi2017} showed that the young classical Cepheids ($<$~140~Myr) are distributed in a geometry that is less elongated than the older ones. Inspection of their Fig.~15 suggests that most of the young Cepheids are located at 60--67.5~kpc in the SMC's central area. This corresponds to a $-$0.07/+0.19~mag difference from the adopted distance modulus. The potential zero-point difference is small with respect to the size of the UMS box; thus, we do not consider the effect of SMC's line-of-sight depth in our analysis.

\subsection{The KDE Map}
\label{kde.sec}

\begin{figure*}[htb!]
\centering
\includegraphics[scale=0.75,angle=0]{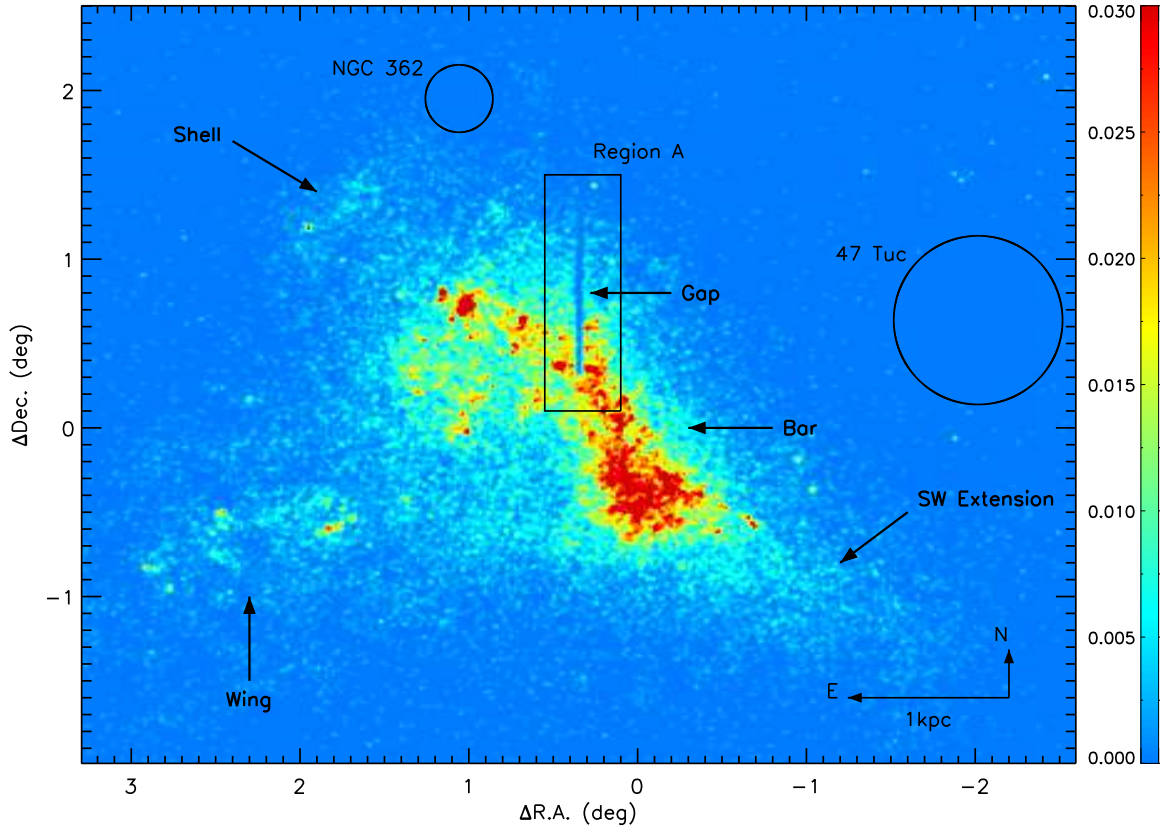}
\caption{KDE map of the UMS stars. The color bar is in units of number of stars~pc$^{-2}$. The two circles show the globular-cluster regions (47~Tuc and NGC~362) which have been excluded from our analysis. The vertical strip devoid of stars at $\Delta$R.A.~$\sim$~0.3$^\circ$ arises from the gap between tiles SMC~5\_3 and 5\_4. The rectangle shows Region~A, which is used to fill the gap with MCPS data (see Section~\ref{mcps.sec}). The image is centered at R.A.(J2000)~=~00$^{\rm h}$50$^{\rm m}$20$^{\rm s}$.8, Dec.(J2000)~=~$-$72$^\circ$49$\arcmin$43$\arcsec$.}
\label{kde.fig}
\end{figure*}

We construct a surface density map of the selected UMS stars using kernel density estimation (KDE). This is done by convolving the stellar distribution with a Gaussian kernel. In principle, the choice of the kernel width is arbitrary and depends on the scale of the object we are interested in. A narrow kernel preserves the small-scale structures in the map but may suffer from large statistical fluctuations. On the other hand, a wide kernel reduces statistical noise, but the resultant KDE map would have poor spatial resolution. In this section, we try to build a KDE map where the small-scale structures are still preserved. As a result, we look for a kernel that is as small as possible. Based on tests with different widths, we found that a Gaussian kernel with a width of 10~pc (standard deviation) is appropriate to balance spatial resolution and noise. With this kernel, it is possible to resolve structures down to scales of $\sim$10~pc. The corresponding KDE map is shown in Fig.~\ref{kde.fig}.

A number of well-known features are evident in the KDE map. The majority of UMS stars are distributed in the main body of the SMC, which is northeast-southwest elongated and also known as the {\it Bar}.  A diffuse {\it Wing} of stars resides to the east of the Bar, and another stellar extension is seen in the extreme SW ({\it SW Extension}). To the northeast there is a diffuse {\it Shell} of stars perpendicular to the Bar. All of these features are labeled in the figure, and they agree well with previous studies by, e.g., \citet{Cioni2000b} and \citet{Zaritsky2000}.

There is a vertical strip devoid of stars at $\Delta$R.A.~$\sim$~0.3$^\circ$, which arises from the gap between VMC tiles SMC~5\_3 and 5\_4 (see Fig.~\ref{map.fig}). However, our map analysis requires a contiguous spatial coverage. Thus, we use data from the MCPS survey \citep{Zaritsky2002, Zaritsky2004} to fill this gap, which is detailed in the following subsection.

\subsection{Filling the Gap with MCPS}
\label{mcps.sec}

\begin{figure}[htb!]
\centering 
\includegraphics[scale=0.6,angle=0]{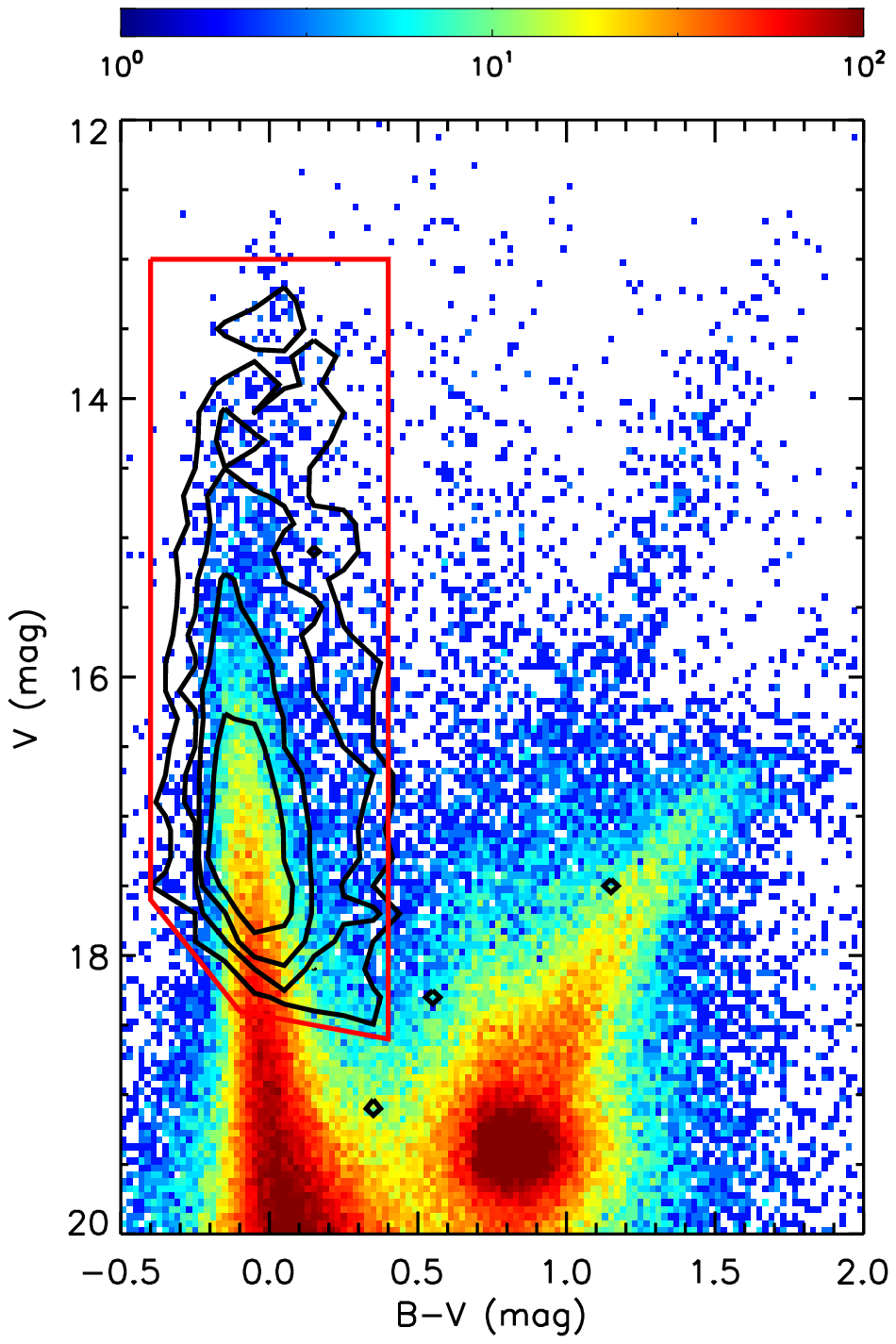}
\caption{($B - V$, $V$) color--magnitude Hess diagram of MCPS sources in Region~A. The color scale shows the number of stars in each color--magnitude bin, with a bin size of 0.02~mag in color and 0.05~mag in magnitude. The black contours show the distribution of MCPS sources in Region~A which are spatially cross-matched with the VMC UMS sample. For better statistics, the contours are computed with star counts using larger bin sizes of 0.1~mag in color and 0.2~mag in magnitude. The contour levels are 0.01, 0.03, 0.1, and 0.3 of the maximum value (303 stars) from the outside to the inside. The red polygon shows the criteria adopted for selecting UMS stars from this CMD.}
\label{mcps.fig}
\end{figure}

The MCPS survey is a photometric survey of the Magellanic Clouds in the $U$, $B$, $V$, and $I$ bands. The limiting magnitude is approximately $V$~=~20 and $I$~=~21~mag, depending on the local crowding of the images. To fill the gap, we first define a rectangular region of 0.45$^\circ$~$\times$~1.40$^\circ$ centered on the gap. The region is shown as the solid rectangle in Fig.~\ref{kde.fig} and is referred to as Region~A hereafter.

Next, we select a sample of UMS stars in Region~A from the MCPS data. Figure~\ref{mcps.fig} shows the ($B - V$, $V$) CMD of the MCPS sources in this region. The MS, RGB, and RC are all clearly seen. We cross-matched the MCPS sources with the UMS stars selected in Section~\ref{ums.sec}, using a matching radius of 1$\arcsec$. The matched sources are shown as the contours in Fig.~\ref{mcps.fig}. Thus, we use the red polygon, which approximately encloses the contours, to select a sample of UMS stars based on MCPS photometry. This ``MCPS UMS sample" contains 13,991 stars in total. Note that extinction in this region is relatively low and is not a matter of concern here \citep[see e.g. Fig.~4 of ][]{Haschke2011}.

\begin{figure}[htb!]
\centering
\includegraphics[scale=0.60,angle=0]{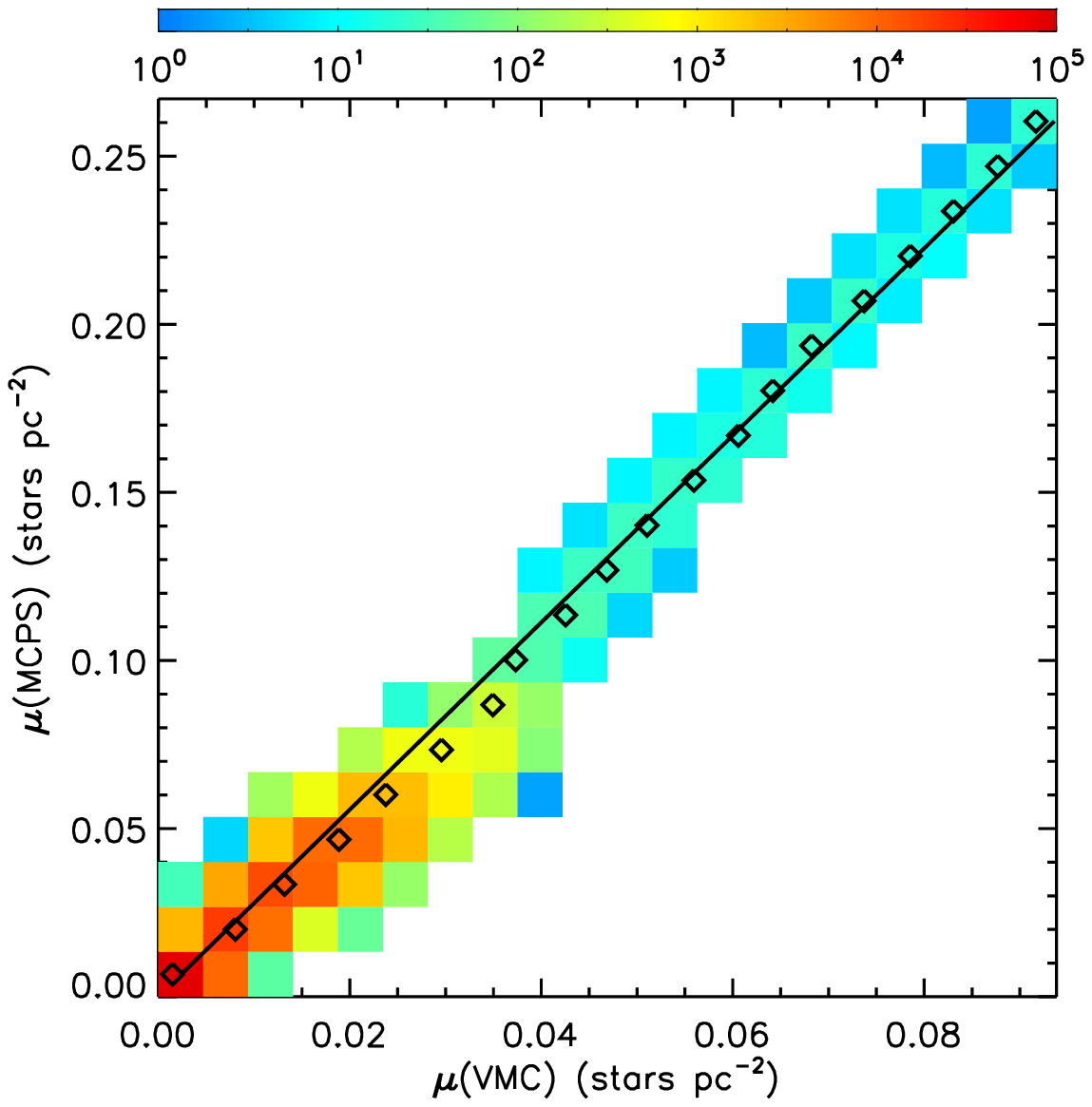}
\caption{Pixel-to-pixel comparison between the VMC map and the MCPS map of Region~A. The comparison diagram is divided into 20~$\times$~20 bins, and the color scale shows the number of pixels in each bin. The points correspond to median values of $\mu_{\rm VMC}$ in equally-spaced bins of $\mu_{\rm MCPS}$. The solid line is a linear function fitted to the points, which is used to calibrate the MCPS map against the VMC one.}
\label{rec.fig}
\end{figure}

\begin{figure}[htb!]
\centering
\begin{tabular}{ccc}
\includegraphics[scale=0.4,angle=0]{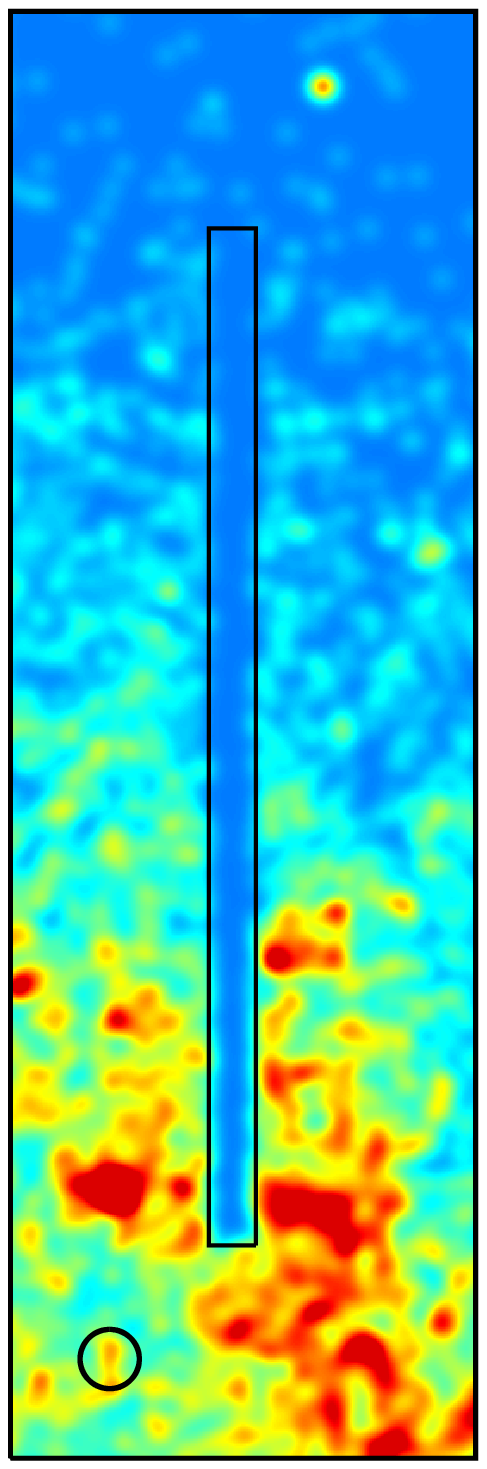}
\includegraphics[scale=0.4,angle=0]{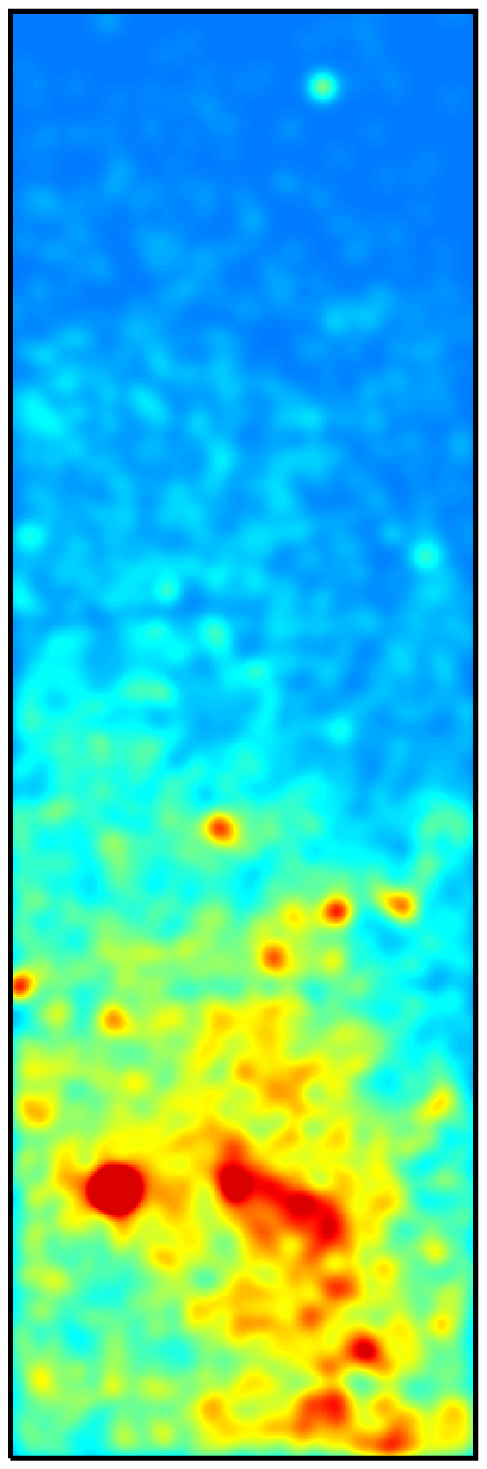}
\includegraphics[scale=0.4,angle=0]{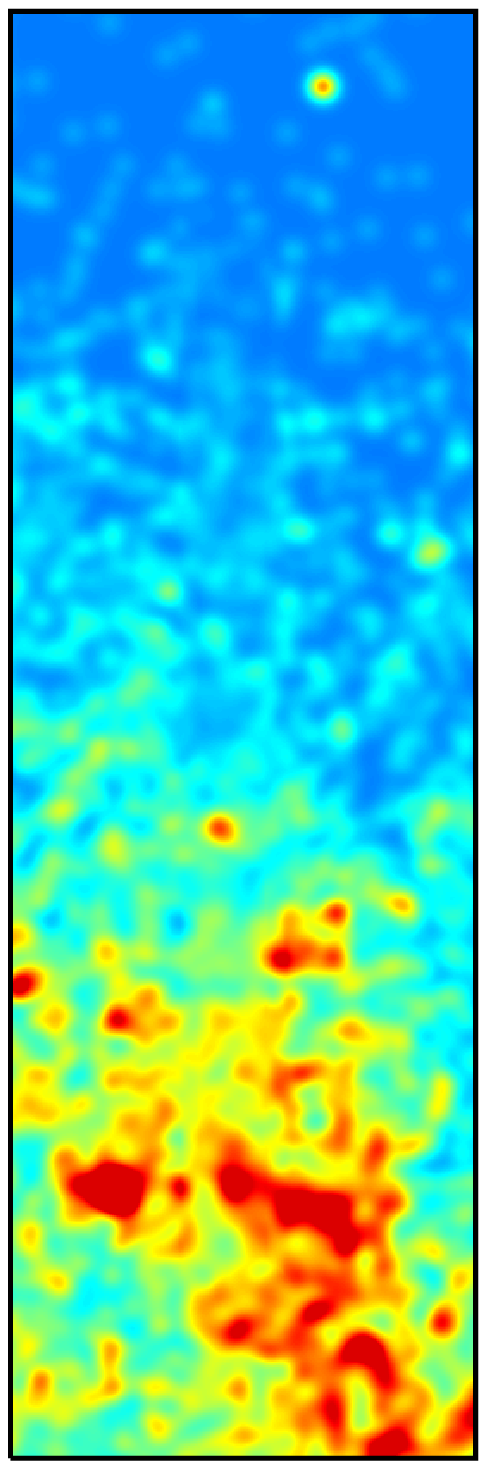}
\end{tabular}
\caption{KDE maps of Region~A. {\it Left-hand panel}: Original map obtained from the VMC UMS sample, a simple zoom-in of Region~A of Fig.~\ref{kde.fig}. {\it Middle panel}: KDE map obtained from the MCPS UMS sample and calibrated against the VMC map. {\it Right-hand panel}: Combined map of VMC and MCPS. In the left-hand panel, the vertical narrow rectangle shows the extent of the gap, and the circle in the bottom left-hand corner has a radius three times the kernel width. The color scales in all three panels are the same as in Fig.~\ref{kde.fig}.}
\label{zoom.fig}
\end{figure}

Third, a KDE map of Region~A is constructed for the MCPS UMS sample in the same way as in the previous subsection. Note that the pixel values of the VMC and MCPS maps are not necessarily the same since they are based on different samples. Figure~\ref{rec.fig} shows the pixel-to-pixel comparison of the VMC and MCPS maps of Region~A. The comparison is done only with the ``good" pixels -- pixels in the gap, within 10~pc (or 34$\arcsec$.4, the kernel width of the KDE maps) from the gap edge, or within 10~pc from the region edge are excluded from the comparison. Figure~\ref{rec.fig} shows that there is a tight correlation between the pixel values of the VMC and MCPS maps. The correlation can be fitted with a linear function (solid line in Fig.~\ref{rec.fig}),
\begin{equation}
\mu_{\rm VMC} =  0.36 \times \mu_{\rm MCPS},
\label{rec.eq}
\end{equation}
where $\mu_{\rm VMC}$ and $\mu_{\rm MCPS}$ are the pixel values of the VMC and MCPS KDE maps, respectively. Thus, we use Eq.~(\ref{rec.eq}) to calibrate the MCPS map against the VMC map. In other words, the two maps can be placed on a uniform scale by assuming that each MCPS-selected UMS star accounts for 0.36 of a VMC-selected one. The original VMC map and the calibrated MCPS map of Region~A are shown in Fig.~\ref{zoom.fig}. In general, they agree with each other very well, except for areas affected by the gap or by the region edge. This ensures that we can fill the gap with MCPS data.

\begin{figure*}[htb!]
\centering
\includegraphics[scale=0.75,angle=0]{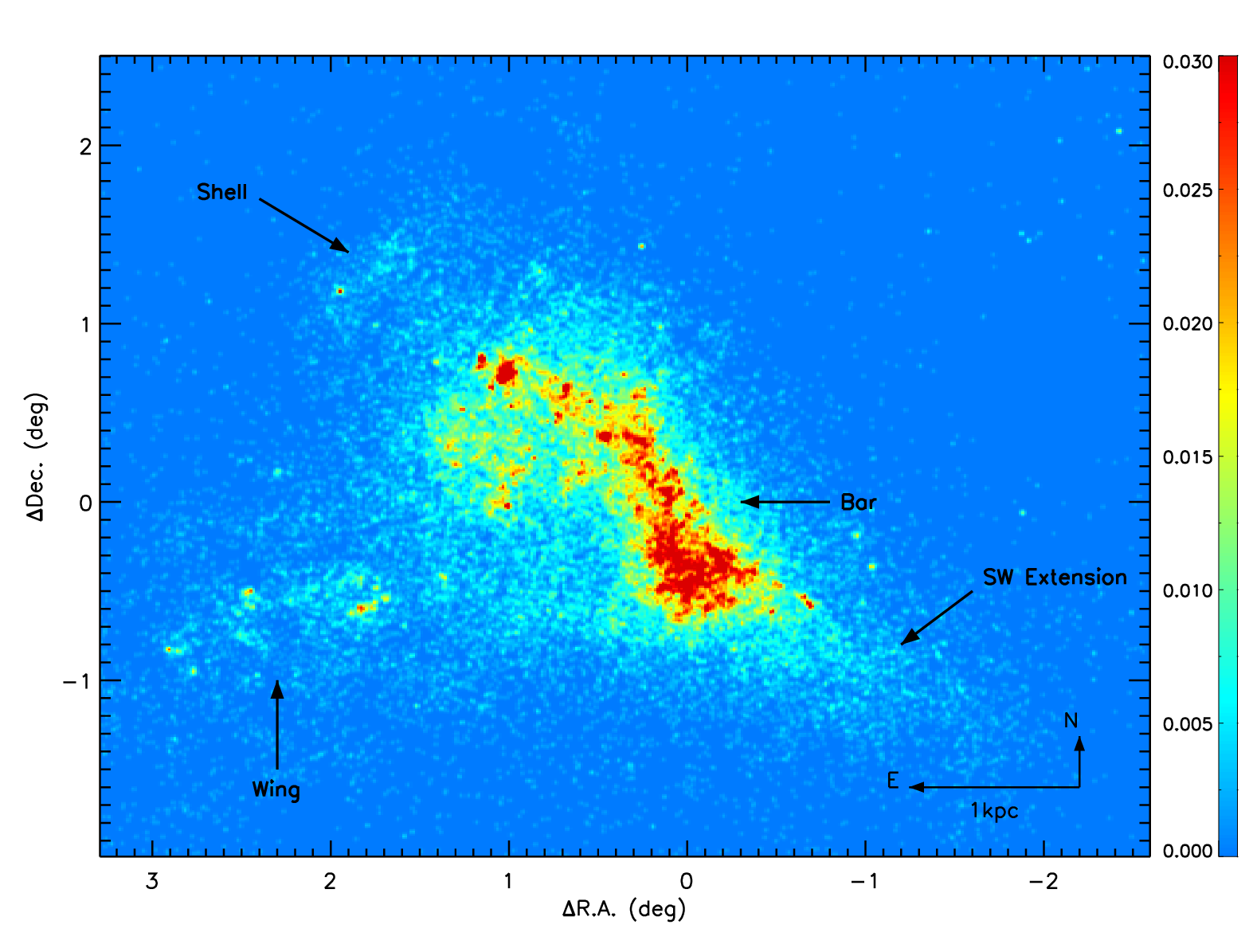}
\caption{As Fig.~\ref{kde.fig} but the gap is filled with MCPS data.}
\label{fix.fig}
\end{figure*}

Finally, we try to combine the two maps. A weight map is constructed, where the pixels have a weight of $w$~=~1.0 if they are more than 30~pc (three kernel widths) from the gap edge; a weight of $w$~=~0.0 is assigned to pixels inside the gap or within 10~pc from the gap edge; linear interpolation is used to assign weights to all other pixels. The VMC and MCPS maps are then combined as
\begin{equation}
\mu_{\rm comb} = \mu_{\rm VMC} \times w + \mu_{\rm MCPS} \times (1 - w),
\label{comb.equ}
\end{equation}
where $\mu_{\rm comb}$ is the pixel value of the combined map. Adopting this method, the combined map (Fig.~\ref{zoom.fig}, right-hand panel), from the outer areas to the gap, changes smoothly from the VMC values to the (calibrated) MCPS values, effectively filling the gap. The complete gap-filled map is shown in Fig.~\ref{fix.fig}. The map has a median value of 2~$\times$~10$^{-5}$~pc$^{-2}$, a mean value of 0.002~pc$^{-2}$, and a standard deviation of 0.004~pc$^{-2}$. Our identification and analysis of the young stellar structures is based on this map.

\section{Identification of Young Stellar Structures}
\label{id.sec}

We use the contour-based map analysis technique proposed by \citet{Gouliermis2015, Gouliermis2017} to identify young stellar structures in the SMC. The idea is to identify them as surface overdensities over a {\it range} of significance levels. In the KDE map (Fig.~\ref{fix.fig}), the iso-density contours are obtained from 1$\sigma$ to 15$\sigma$ in equal steps of 1$\sigma$ ($\sigma$~=~0.004~pc$^{-2}$; note that the median value of the KDE map is very close to zero compared with the standard deviation). On each significance level, any iso-density contour enclosing an overdensity is regarded as a candidate young stellar structure. The iso-density contours are thus regarded as the (projected) ``boundaries" of the candidate young stellar structures.

We next determine physical parameters of the candidates. The size of a structure ($R$) is estimated with the radius based on a circle of the same area as that of its boundary. The structure's mass is characterized by the number of UMS stars inside its boundary ($N_\ast$). To first-order approximation, $N_\ast$ is proportional to the mass of the structure, assuming the structures have similar ages and that the stellar initial mass function (IMF) is fully sampled (\citealt{Gouliermis2015, Sun2017b}; see also Section~\ref{corr.sec} for a more detailed discussion). $N_\ast$ is obtained from simple star counts of the VMC UMS sample. For the gap of VMC observations, we use instead the MCPS UMS sample (according to Eq.~\ref{rec.eq}, each MCPS-selected UMS star accounts for 0.36 of a VMC-selected one). We will hereafter refer to $N_\ast$ as the ``stellar number" for simplicity. Finally, the structure's surface density is calculated as $\Sigma$~=~$N_\ast$/($\pi R^2$).

However, the candidates may include spurious detections as well. A commonly adopted method to reject unreliable detections is to require a minimum stellar number ($N_{\rm min}$) to define a young stellar structure \citep[e.g.][]{Bastian2007, Bastian2009, Gouliermis2015, Gouliermis2017, Sun2017a, Sun2017b}. In other words, a candidate is regarded a reliable young stellar structure only if $N_\ast$~$\ge$~$N_{\rm min}$. The choice of $N_{\rm min}$ is arbitrary -- a large $N_{\rm min}$ can reject most unreliable detections but may remove genuine structures as well; a small $N_{\rm min}$, on the other hand, leads to more unreliable structures. In the following analysis, we adopt $N_{\rm min}$~=~5~stars, which has been used extensively in previous works. Furthermore, candidates on the lowest significance levels may arise from random density fluctuations. However, we cannot rule out the possibility that they may be genuine structures with low surface densities. In order to reject the less probable candidates of this kind, we regard candidates on the 1$\sigma$ and 2$\sigma$ significance levels as genuine structures if their boundaries enclose one or more structures on the 3$\sigma$--15$\sigma$ significance levels.

\begin{figure*}[htb!]
\centering
\includegraphics[scale=0.7,angle=0]{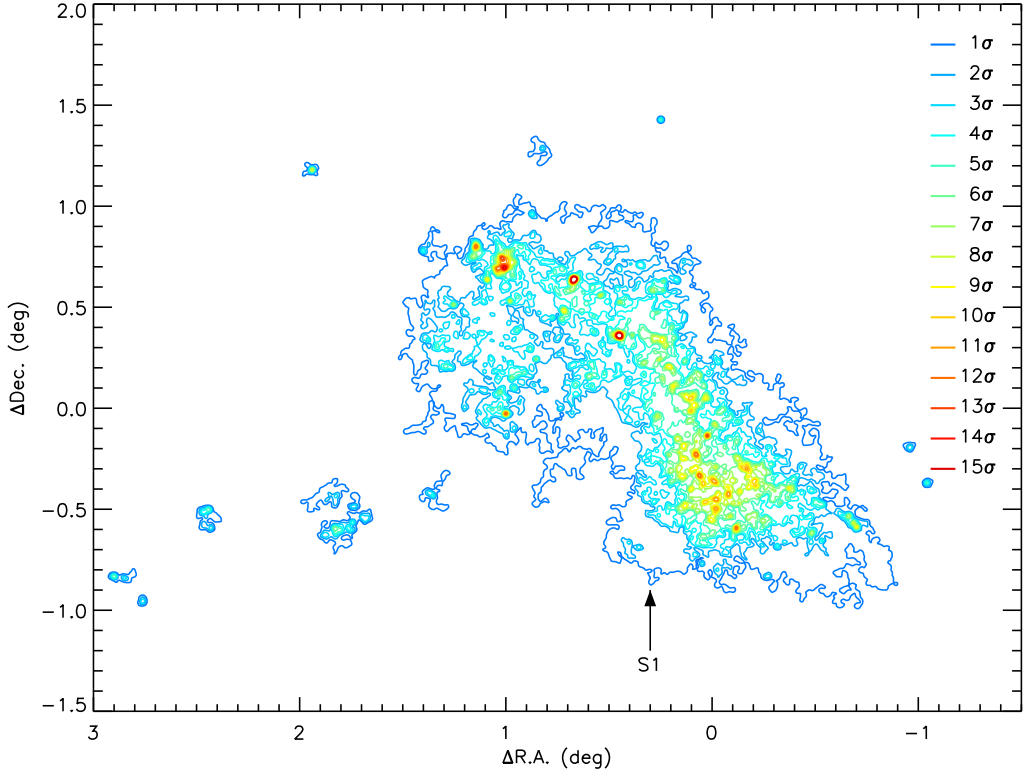}
\caption{Boundaries of all identified young stellar structures, with the colors encoded according to their significance levels. S1, whose dendrogram is shown in Fig.~\ref{tree.fig}, is also labeled.}
\label{struct.fig}
\end{figure*}

After rejecting the less reliable candidates, we finally identify 556 young stellar structures. They include 14, 30, 119, 90, 72, 59, 47, 45, 36, 19, 14, 10, 5, 3, and 3 structures on the 1$\sigma$--15$\sigma$ significance levels, respectively. The structures are listed in Table~1, along with their IDs, significance levels, coordinates, and physical parameters. In the following, we will refer to a specific structure using its ID number with a prefix ``S" (short for ``structure"; for instance, ``S1" refers to the first structure in Table~1). The boundaries of all identified structures are shown in Fig.~\ref{struct.fig}. Most of the identified structures belong to the Bar region, while another group of structures are from the Wing. Very few structures are identified in the Shell or in the SW extension, since the stellar surface density is very low there.

\begin{deluxetable}{rcccrrr}
\tablecolumns{7}
\tablecaption{Identified Young Stellar Structures.\label{catalog.tab}}
\tablehead{
\colhead{ID} & \colhead{level} & \colhead{$\alpha$(J2000)} & \colhead{$\delta$(J2000)} & \colhead{$R$} &  \colhead{$N_\ast$} & \colhead{$\Sigma$}\\
\colhead{ } & \colhead{($\sigma$)} & \colhead{(deg)} & \colhead{(deg)} & \colhead{(pc)} & \colhead{ } & \colhead{(pc$^{-2}$)} \\
\colhead{(1)} & \colhead{(2)} & \colhead{(3)} & \colhead{(4)} & \colhead{(5)} & \colhead{(6)} & \colhead{(7)}}
\startdata
   1 &  1 &  13.6355 & $-$72.7921 &1046.2 &31,487 & 0.0092 \\
   2 &  1 &  18.8066 & $-$73.3464 & 110.5 &  269 & 0.0070 \\
   3 &  1 &  19.0798 & $-$73.1572 &  87.6 &  125 & 0.0052 \\
   4 &  1 &  21.0529 & $-$73.2027 &  64.0 &  106 & 0.0082 \\
   5 &  1 &  17.2412 & $-$73.1845 &  82.2 &   87 & 0.0041 \\
   6 &  1 &  15.2237 & $-$71.5373 &  61.3 &   61 & 0.0052 \\
   7 &  1 &  22.6459 & $-$73.4161 &  48.9 &   53 & 0.0070 \\
   8 &  1 &  18.6065 & $-$73.2029 &  37.9 &   32 & 0.0071 \\
   9 &  1 &  18.7448 & $-$71.5467 &  38.4 &   42 & 0.0091 \\
  10 &  1 &   9.3148 & $-$72.9954 &  26.8 &   22 & 0.0098 \\
  11 &  1 &  22.3801 & $-$73.5534 &  26.0 &   20 & 0.0094 \\
  12 &  1 &   8.9750 & $-$73.1664 &  24.6 &   21 & 0.0110 \\
  13 &  1 &  13.3652 & $-$71.3993 &  18.4 &   14 & 0.0131 \\
  14 &  1 &   5.2641 & $-$70.6099 &  15.7 &    9 & 0.0116 \\
  15 &  2 &  13.9915 & $-$72.7453 & 836.9 &24,288 & 0.0110
\enddata
\tablecomments{{\it Column~1}: ID number for each young stellar structure; {\it Column~2}: significance level; {\it Column~3}: right ascension of the center of each young stellar structure, defined as $\alpha$~=~($\alpha_{\rm min}$~+~$\alpha_{\rm max}$)/2, where $\alpha_{\rm min}$ and $\alpha_{\rm max}$ are the minimum and maximum right ascension of the iso-density contour of each structure, respectively; {\it Column~4}: same as Column~3 but for the declination; {\it Columns~5--7}: size, stellar number, and surface density. Only the first 15 records are shown as an example. The complete catalog is available online.}
\end{deluxetable}

Also note that some (but not all) structures on low significance levels correspond to significant portions of the whole SMC. In fact, the largest structure S1 covers essentially the whole main body of the galaxy. Only on higher significance levels do they break down to independent, internal structures on sub-galactic scales. This is very different from disk galaxies, where the young stellar structures are clearly distinct even on low significance levels \citep[e.g.][]{Gouliermis2015, Gouliermis2017}. This is possibly because disk galaxies have a more ordered structure, while the SMC is more irregular and its stellar populations are more mixed.

\section{Properties of Young Stellar Structures}
\label{property.sec}

\subsection{The Hierarchical Pattern}
\label{tree.sec}

\begin{figure*}[htb!]
\centering
\includegraphics[scale=0.72,angle=0]{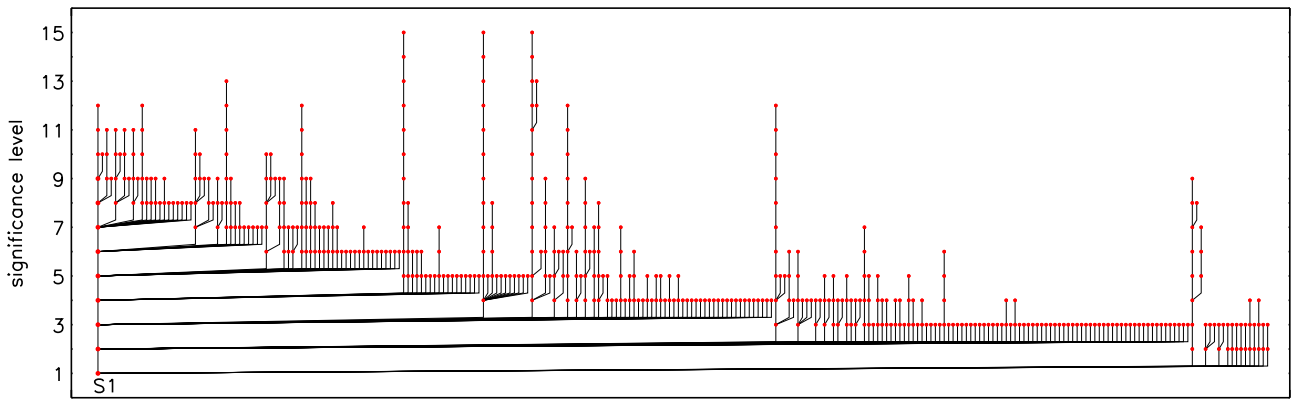}
\caption{Dendrogram of the young stellar structures. The red points indicate the structures on different significance levels, and the black links show their ``parent--child" relations. The root corresponds to S1 on the 1$\sigma$ significance level, and the branches and leaves represent its substructures on higher significance levels.}
\label{tree.fig}
\end{figure*}

Figure~\ref{struct.fig} shows that the young stellar structures are organized in a hierarchical manner. Generally, a structure on a low significance level may contain one or more substructures on the higher significance levels within its boundary. To illustrate this behavior intuitively, we show in Fig.~\ref{tree.fig} the dendrogram of structure S1 on the 1$\sigma$ significance level. Dendrograms are structure trees showing the ``parent--child" relations of young stellar structures on various significance levels. In S1's dendrogram, the ``root" corresponds to S1 itself, while the ``branches" and ``leaves" represent its substructures on the 2$\sigma$--15$\sigma$ significance levels. It is obvious that S1 branches into a lot of substructures on the 2$\sigma$ significance level, many of which in turn branches into substructures on the 3$\sigma$ significance level. This fragmentation-like behavior continues to higher significance levels until the top (15$\sigma$) significance level. Thus, the young stars in the SMC exhibit a high degree of hierarchical subclustering, which is typical in many star-forming regions and galaxies \citep[e.g.][]{Gouliermis2010, Gouliermis2015, Gusev2014, Sun2017a, Sun2017b}.

\subsection{Morphological Irregularity}
\label{morph.sec}

\begin{figure}[htb!]
\centering
\includegraphics[scale=0.75,angle=0]{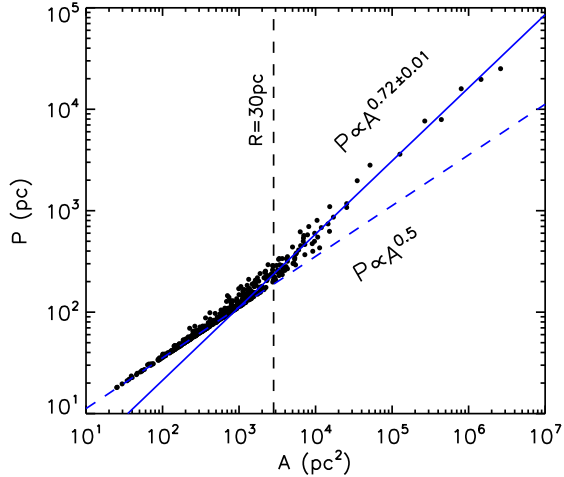}
\caption{Perimeter-area relation of the boundaries of all identified young stellar structures. The blue dashed line corresponds to that of geometric circles. The vertical dashed line corresponds to $R$~=~30~pc (according to the definition of structure size). Beyond this limit the data points are fitted with a power law, as shown by the blue solid line.}
\label{peri.fig}
\end{figure}

Figure~\ref{struct.fig} also shows that many of the young stellar structures have very irregular boundaries, which are significantly different from smooth curves. This morphological irregularity can be characterized by the perimeter--area relation. In Fig.~\ref{peri.fig} we show the perimeters ($P$) and areas ($A$) of the boundaries of all identified young stellar structures. In the plot, the blue dashed line shows the perimeter--area relation of geometric circles, which is characterized by a power-law slope of $\alpha_p$~=~0.5. This relation ``clips" the data points at $A$~$<$~3~$\times$~10$^3$~pc$^2$ by coinciding with their lower limit. Data points at $A$~$<$~3~$\times$~10$^2$~pc$^2$ are highly consistent with this relation. This reflects the resolution effect for the small young stellar structures. Since the KDE map (Fig.~\ref{fix.fig}) is based on a kernel width of 10~pc, structures smaller than or comparable to this size will appear nearly roundish because of the kernel smoothing.

Beyond $A$~=~3~$\times$~10$^3$~pc$^2$, the data points seem not to be strongly ``clipped" by the perimeter--area relation of geometric circles, suggesting that resolution effects are no longer important. Note, according to the definition of structure size (Section~\ref{id.sec}), this limit corresponds to $R$~=~30~pc, or three times the map resolution\footnote{As will be shown in Section~\ref{dist.sec}, the identification of young stellar structures is complete beyond 10~pc (the KDE map resolution). However, we caution that the perimeter--area relation between 10--30~pc may still be affected by the resolution effect. This is because a structure can be detected as long as its size is beyond the map resolution (and $N_\ast \ge N_{\rm min}$ in order not to be rejected), but the shape of its boundary can only be revealed when its size is much larger than the resolution. Otherwise, its shape will still appear roundish due to kernel smoothing. In other words, for a given structure, one needs a lower resolution to detect it, but a higher resolution to resolve its shape.}. The young stellar structures have a power-law perimeter--area relation beyond this limit. The power-law slope is $\alpha_p$~=~0.72~$\pm$~0.01 based on a least-squares fit. The perimeter--area dimension, $D_p$, is defined such that \textbf{\color{black} $P \propto A^{D_p/2}$} \citep{Falgarone1991}. Thus, the boundaries have a perimeter--area dimension of $D_p~=~1.44~\pm~0.02$. This is significantly larger than their topological dimension of unity, which quantitively reflects the morphological irregularity of the young stellar structures.

\subsection{Mass--Size and Surface Density--Size Relations}
\label{corr.sec}

\begin{figure}[htb!]
\centering
\begin{tabular}{c}
\includegraphics[scale=0.75,angle=0]{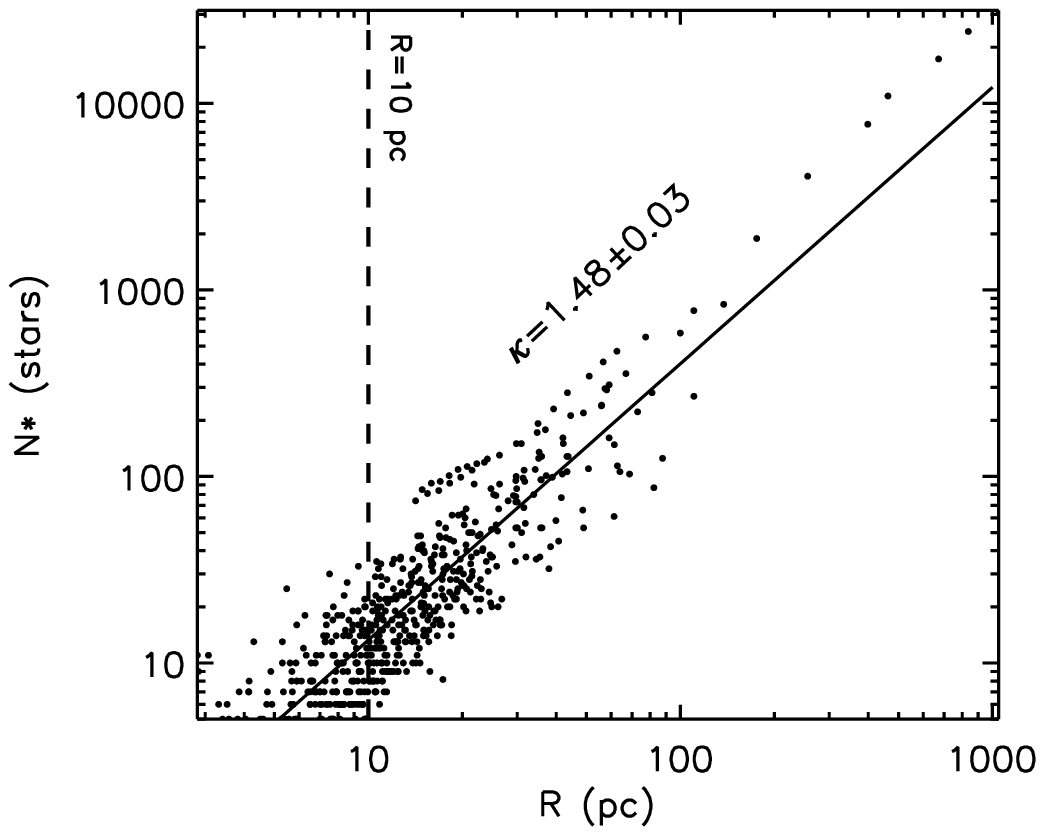} \\
\includegraphics[scale=0.75,angle=0]{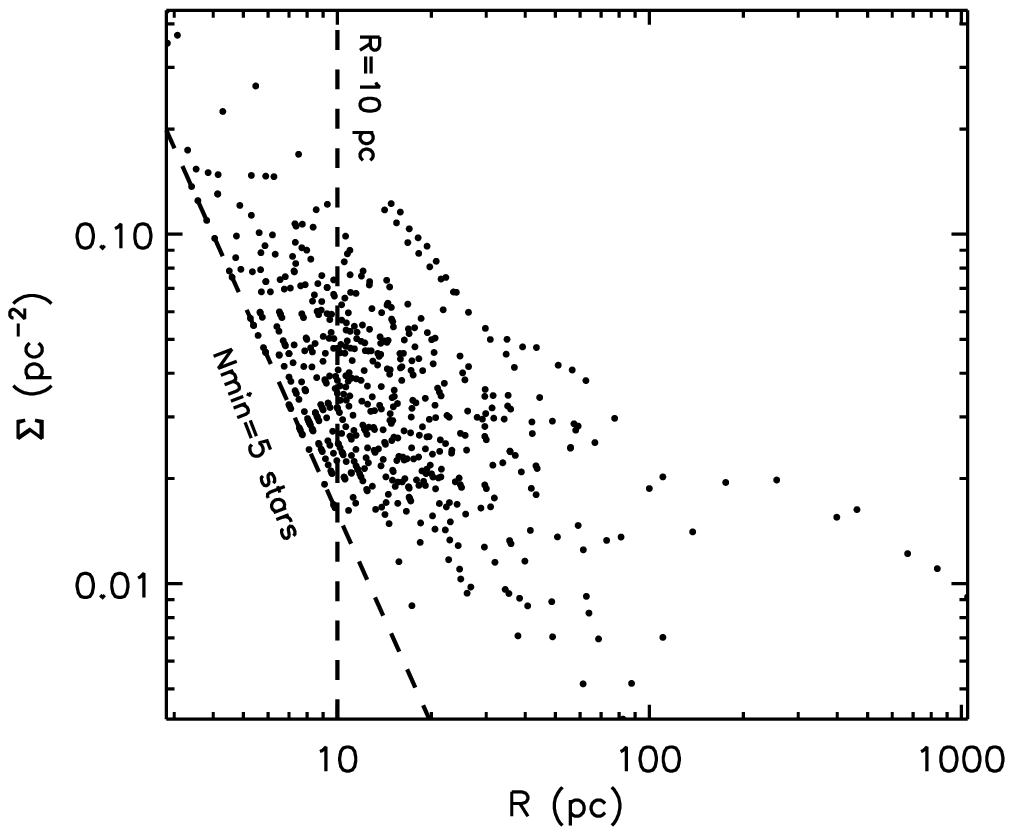}
\end{tabular}
\caption{Stellar number--size (top) and surface density--size (bottom) diagrams of all identified young stellar structures. The dashed lines correspond to the spatial resolution (10~pc) of the KDE map (Fig.~\ref{fix.fig}) and the minimum stellar number ($N_{\rm min}$~=~5~stars) for young stellar structures.}
\label{corr.fig}
\end{figure}

Figure~\ref{corr.fig} shows the stellar number--size and surface density--size diagrams of all identified young stellar structures. With a Pearson correlation coefficient of $-$0.59, the surface densities and sizes of the structures do not show any significant correlations. In contrast, the young stellar structures exhibit a tight power-law correlation between their stellar numbers and sizes, with a Pearson correlation coefficient of 0.90. Using a least-squares fit (solid line) we find that the best-fitting slope is $\kappa~=~1.48~\pm~0.03$.

\begin{figure}[htb!]
\centering
\includegraphics[scale=0.75,angle=0]{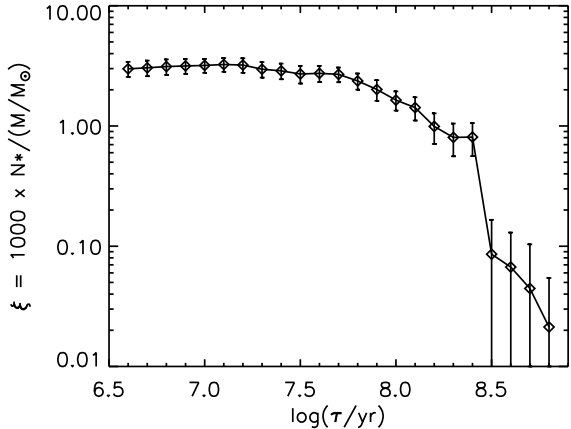}
\caption{Stellar number-to-mass ratios of simulated stellar populations of different ages. The points correspond to the median values of 100 realizations for each age, while the error bars reflect their standard deviations.}
\label{ratio.fig}
\end{figure}

As mentioned in Section~\ref{id.sec}, the stellar number of a structure is proportional to its mass, if the stellar IMF is fully sampled and if the structures have similar ages. This approximation will be acceptable if the stellar number-to-mass ratio ($\xi$) does not change significantly among the young stellar structures. To quantitively assess whether this is valid, we carry out numerical simulations of stellar populations with 20\% solar metallicity \citep[typical of young stars in the SMC;][]{Russell1992} and with ages from log($\tau$/yr)~=~6.6 to 8.8 in steps of 0.1. For each population, a total of 20,000 stars are sampled randomly from the \citet{Chabrier2001} lognormal IMF. We then assume that 30\% of them are unresolved binary systems \citep{Elson1998, Li2013}, and assign masses to their secondary stars based on a flat distribution of primary/secondary mass ratios from 0.7 to 1.0 (below this limit the secondary star has little influence on the system's overall brightness). This corresponds to an initial mass of $\sim$1.55~$\times$~10$^4$~M$_\odot$ in total for each stellar population. The $J$ and $K_s$ magnitudes for the simulated stars are derived using PARSEC isochrones and shifted to SMC's distance modulus. The UMS stars of each stellar population are selected based on the same criterion as used for the real data (Section~\ref{ums.sec}), and the stellar number-to-mass ratios are then calculated as $\xi = 1000 \times N_\ast /(M/M_\odot)$.

We carry out 100 realizations for each age. The mean values of their stellar number-to-mass ratios are shown as the data points in Fig.~\ref{ratio.fig}. Starting from $\xi \sim 3 $ at log($\tau$/yr)~=~6.6, $\xi$ slightly increases to a maximum of 3.3 at log($\tau$/yr)~=~7.1--7.2, and then declines gradually to 1.7 at log($\tau$/yr)~=~8.0, until it reaches 0.8 at log($\tau$/yr)~=~8.4. Within this age limit, $\xi$ differs by at most a factor of $\sim$4. $\xi$ drops sharply at log($\tau$/yr)~=~8.5 by an order of magnitude and continues to decrease toward older ages. Stellar populations older than log($\tau$/yr)~=~8.8 have no or negligible numbers of stars that meet the UMS selection criteria adopted in this work.

As discussed in Section~\ref{ums.sec}, statistically speaking most SMC young stellar structures should have ages younger than or comparable to $\sim$100~Myr. For these structures, the stellar number-to-mass ratio is almost constant around 2, with minimum and maximum values of 1.7 and 3.3, respectively. Structures with ages log($\tau$/yr)~=~8.0--8.4 ($\sim$100--250~Myr) have lower stellar number-to-mass ratios, but they are still not very different. In Section~\ref{ums.sec} we have mentioned that 96\% of stars in the UMS sample are younger than 250~Myr. Thus, we expect very few structures, if any, older than this age. Based on this analysis, we suggest that the variation of $\xi$ caused by evolutionary effects should be small for most SMC young stellar structures.

On the other hand, the error bars in Fig.~\ref{ratio.fig} show the standard deviations of $\xi$ of the 100 realizations for each age. They reflect the variation of $\xi$ caused by the stochastic sampling of the stellar IMF. This variation is also very small for structures younger than log($\tau$/yr)~=~8.4. Note that the simulated populations younger than this age have typically 10--50 UMS stars. Most observed structures contain a few tens of UMS stars, and some large ones even contain hundreds or thousands of UMS stars. Thus, we conclude that the variation of $\xi$ caused by stochastic sampling is also unimportant.

From the above analysis, we suggest that the stellar number is proportional to the mass for most of the young stellar structures. As a result, the stellar number--size relation can be regarded as the mass--size relation with approximately the same slope. The mass--size relation of substructures in a fractal follows $M \propto R^D$, where $D$ is, by definition, the fractal dimension\citep{Mandelbrot1983}. Thus, the young stellar structures have a {\it projected} fractal dimension of $D_2$~=~$\kappa$~=~1.48~$\pm$~0.03. We use the subscript ``2" to emphasize that the young stellar structures are identified from a two-dimensional KDE map of the projected stellar distributions (Section~\ref{id.sec}). $D_2$ reflects significant lumpiness of the young stellar structures, since it is much smaller than 2, the latter being expected for a smooth spatial distribution.

We also investigated the mass--size relations for subsamples of young stellar structures at different significance levels. The corresponding power-law slopes, obtained via least-squares fits, are displayed in Table~\ref{indices.tab}. We find that the mass--size relation has a dependence on the significance levels, since the 1--2$\sigma$ structures have steeper slopes than those on higher significance levels. This behavior is similar to that found by \citet{Gouliermis2017} for the spiral galaxy NGC~1566.

\begin{deluxetable}{cc}
\tablecolumns{2}
\tablecaption{Power-law slopes of mass--size relations.\label{indices.tab}}
\tablehead{
\colhead{level ($\sigma$)} & \colhead{$\kappa$}}
\startdata
   1  &  1.88~$\pm$~0.09  \\
   2  &  1.77~$\pm$~0.06  \\
   3  &  1.65~$\pm$~0.04  \\
   4  &  1.66~$\pm$~0.04  \\
   5  &  1.58~$\pm$~0.06  \\
   6  &  1.64~$\pm$~0.05  \\
   7  &  1.61~$\pm$~0.06  \\
   8  &  1.48~$\pm$~0.06  \\
   9  &  1.60~$\pm$~0.08  \\
  10  &  1.7~$\pm$~0.1  \\
  11  &  1.7~$\pm$~0.2  \\
  12  &  1.6~$\pm$~0.2  \\
$\ge$13  &  1.6~$\pm$~0.2  \\
total & 1.48~$\pm$~0.03
\enddata
\end{deluxetable}

\subsection{Size, Mass, and Surface Density Distributions}
\label{dist.sec}

\begin{figure}[htb!]
\centering
\begin{tabular}{c}
\includegraphics[scale=0.75,angle=0]{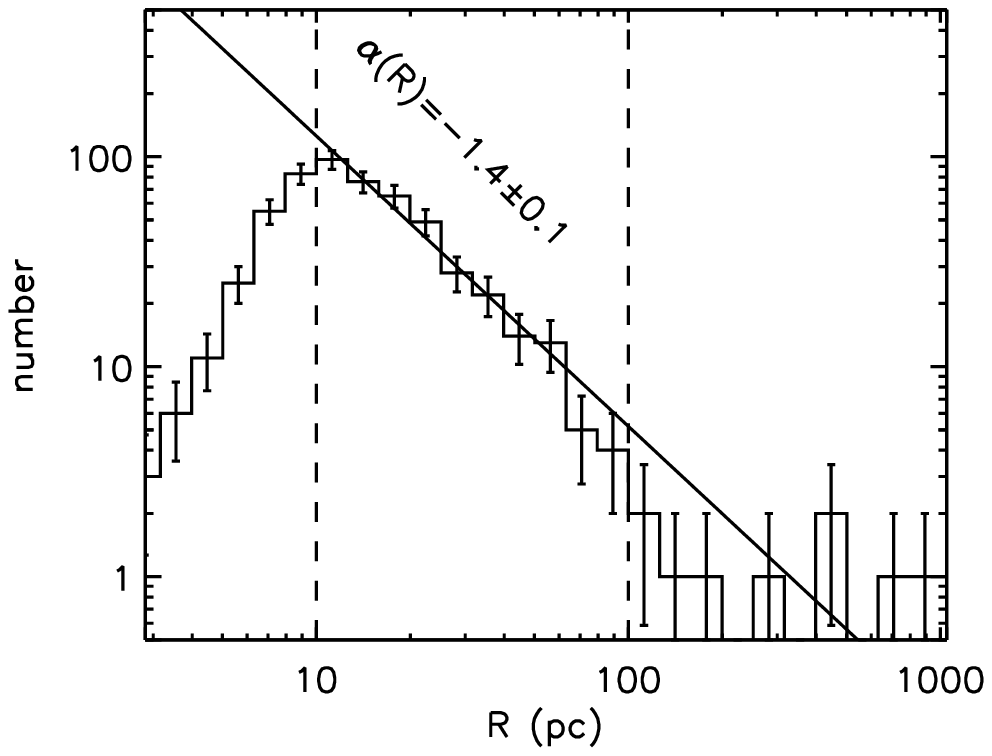} \\
\includegraphics[scale=0.75,angle=0]{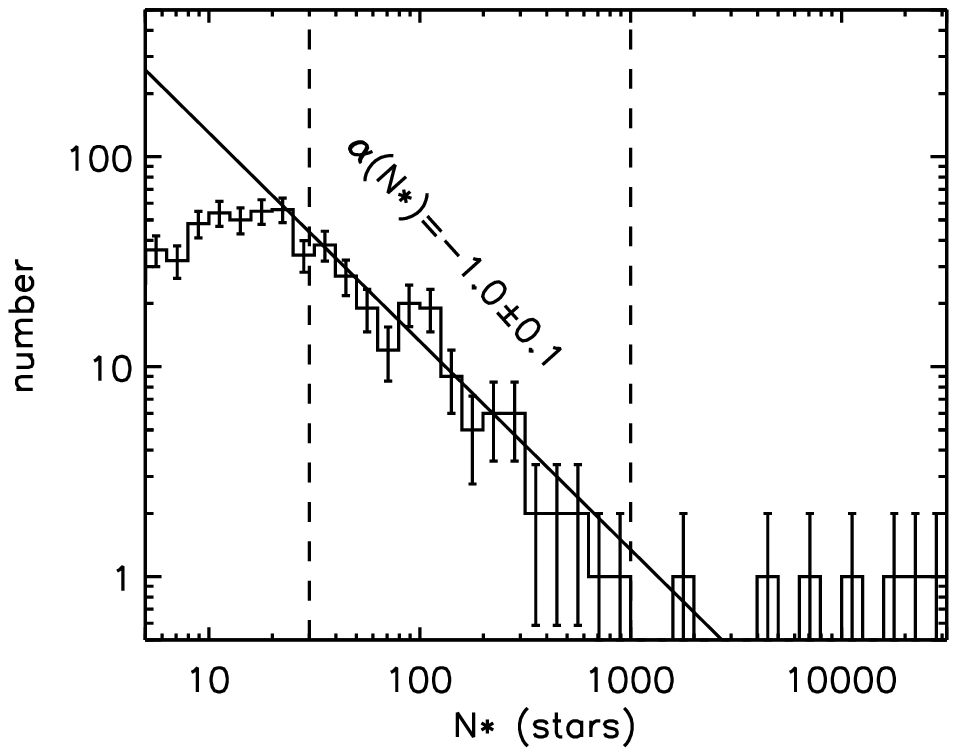} \\
\includegraphics[scale=0.75,angle=0]{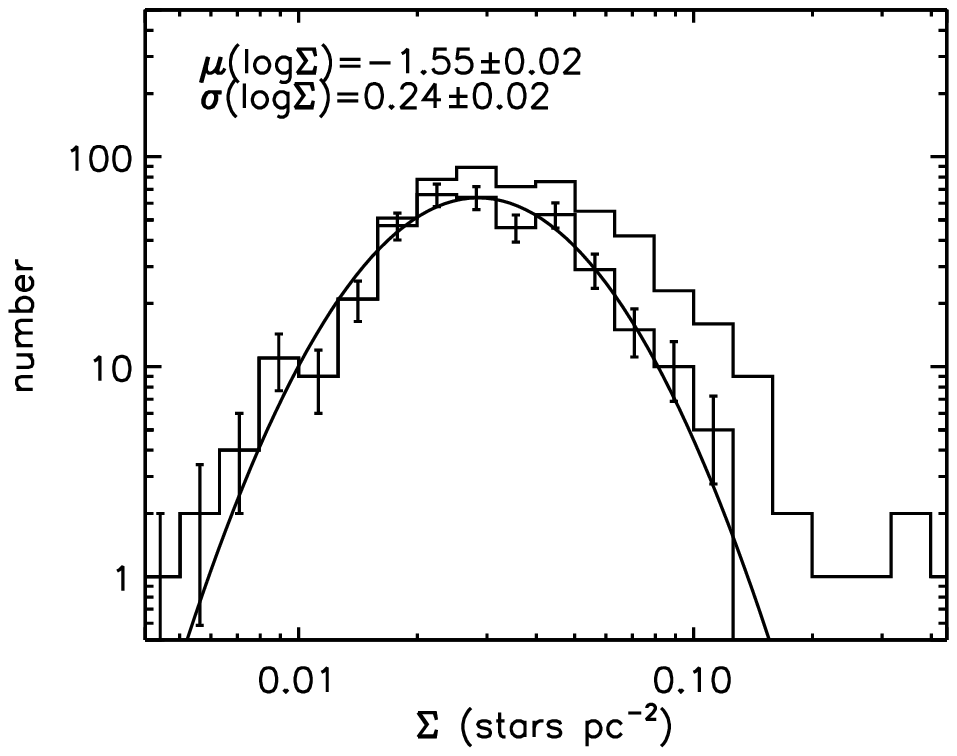}
\end{tabular}
\caption{Size (top), stellar number (middle), and surface density (bottom) distributions of the young stellar structures. The distributions are calculated per logarithmic interval, and have a bin size of 0.1~dex, respectively. In the top and middle panels, the solid lines show power-law fits to the distributions, and the two vertical dashed lines show the fitting intervals. In the bottom panel, the histogram without error bars corresponds to all identified young stellar structures, while the histogram with error bars corresponds to structures larger than $R~=~10$~pc. The two histograms overlap at small $\Sigma$, and the latter one is fitted with a lognormal distribution, as shown by the solid-line curve. In all three panels, the error bars reflect Poissonian uncertainties.}
\label{dist.fig}
\end{figure}

Figure~\ref{dist.fig}~(top panel) shows the size distribution of all identified young stellar structures. The distribution peaks at $\sim$10~pc, with a drop toward small sizes. This drop is caused at least in part by the incompleteness of young stellar structures \citep[see also][]{Sun2017a, Sun2017b}. On the one hand, the KDE map (Fig.~\ref{fix.fig}), from which the structures are identified, has a spatial resolution of 10~pc (Section~\ref{map.sec}). As a result, structures smaller than 10~pc are incomplete due to the finite spatial resolution. On the other hand, we have arbitrarily required each structure to contain at least $N_{\rm min}$~=~5~stars (Section~\ref{id.sec}). Although this criterion can reject less reliable detections, it may also remove some genuine structures. This effect can also be assessed with the stellar number--size and surface density--size diagrams (Fig.~\ref{corr.fig}). In the diagrams, the spatial resolution and minimum stellar number are indicated by the dashed lines. It is obvious that structures at small $R$ are incomplete, especially for those with low surface densities.

The size distribution becomes increasingly noisy toward large sizes. In the range of 10--100~pc the size distribution is consistent with a single power law. Thus, we have performed such a fit to the data, which is shown as the solid line in Fig.~\ref{dist.fig}~(top panel). The best fitting slope is $\alpha(R)~=~-1.4~\pm~0.1$. \textbf{Substructures inside a fractal follow a cumulative size distribution
\begin{equation}
N(>R) \propto R^{-D},
\label{frac1.eq}
\end{equation}
where $D$ is the fractal dimension \citep{Mandelbrot1983}. Because Eq.~\ref{frac1.eq} is a single power law, it is mathematically equivalent to a differential size distribution (per logarithmic interval, as in Fig.~\ref{dist.fig})
\begin{equation}
{\rm d}N/{\rm dlog} R \propto R^{-D}.
\label{frac2.eq}
\end{equation}}
Equation~\ref{frac2.eq} has the same form as the size distribution of young stellar structures with 10~$<$~$R$~$<$~100~pc. Thus, these structures agree with a fractal dimension of $D_2$~=~$-\alpha(R)$~=~1.4~$\pm$~0.1, which is very close to that derived from the mass--size relation (1.48~$\pm$~0.03; Section~\ref{corr.sec}).

At $R$~=~300~pc the power law drops below 1, but there are still a few structures larger than this size. We have checked the boundaries of these structures, whose linear scales are found to be comparable to that of the SMC's main body\footnote{Note $R$ is defined with the area enclosed by the structure's boundary; elongated structures may have linear scales significantly larger than $R$.}. As mentioned in Section~\ref{id.sec}, they trace the extent of the SMC's galactic structure rather than internal structures on sub-galactic scales. We will discuss later (Section~\ref{hierarchical.sec}) that the formation of the internal, sub-galactic structures are possibly controlled by supersonic turbulence in the ISM. In contrast, those larger structures are obviously more influenced by global galactic processes. As a result, they do not follow the same power-law size distribution as the sub-galactic ones. Given their small number, we do not attempt to investigate their size distribution in greater detail.

Similar to the size distribution, the stellar number distribution (Fig.~\ref{dist.fig}, middle panel) is also subject to incompleteness at small values. This is evident in the stellar number--size diagram (Fig.~\ref{corr.fig}~top), which clearly shows that the spatial resolution of 10~pc rejects many small-$N_\ast$ structures. This effect is no longer important for structures with $N_\ast~>~30$~stars. Thus, we have fitted the distribution in the range of 30--1000~stars with a single power law. The best-fitting result is shown in Fig.~\ref{dist.fig} (middle panel), with a slope of $\alpha(N_\ast)~=~-1.0~\pm~0.1$. As demonstrated in the previous subsection, $N_\ast$ is proportional to structure mass. Thus, the stellar number distribution reflects the structures' mass distribution, which is also a power law with nearly the same slope. It can be derived mathematically that the mass distribution (per logarithmic interval) of substructures inside a fractal should have a power-law slope of $-$1, irrespective of the fractal dimension \citep[e.g.][their Section~4]{Elmegreen1996}. This is in agreement with the measurement obtained here.

The surface density distribution is shown in the bottom panel of Fig.~\ref{dist.fig}. However, it is not easy to assess the influence of incompleteness, since surface density is not correlated with size (Fig.~\ref{corr.fig}, bottom panel) -- the effects of spatial resolution and minimum stellar number reject structures of a wide range of surface densities, especially at small sizes and small surface densities. Based on the large scatter of the surface density--size relation, we assume that structures larger than 10~pc should be representative of their underlying young stellar structure population in terms of surface density distributions. Their surface density distribution is shown in Fig.~\ref{dist.fig} (bottom panel) as the histogram with error bars. The distribution is in good agreement with the solid-line curve, which is a lognormal function fitted to the data.

The appearance of a histogram may depend sensitively on the bin size \citep{Ivezic2014}. We have tested this effect by changing the bin sizes of the parameter distributions. The power-law slopes of the size/stellar number distributions and the lognormal form of the surface density distribution do not show any significant deviations. Thus, the results in this subsection are robust against this effect.

\section{Discussion}
\label{discussion.sec}

\subsection{Imprints from turbulent ISM}
\label{hierarchical.sec}

There are noticeable similarities between the young stellar structures and the ISM. Similarly to the hierarchical young stellar structures, the ISM also contains significant substructures in a hierarchical manner, including clouds, clumps, and cores on all scales. The ISM substructures exhibit highly irregular morphologies. The projected boundaries of molecular clouds have been investigated based on the perimeter--area relation in a number of studies \citep[e.g.,][]{Beech1987, Scalo1990, Falgarone1991, Vogelaar1994, Lee2016}. The typical perimeter--area dimensions are 1.4--1.5, very close to $D_p~=~1.44~\pm~0.02$ as derived in Section~\ref{morph.sec} for the young stellar structures \citep[although smaller values have also been reported for several molecular clouds; see e.g.][]{Dickman1990, Hetem1993}. Also similar to the young stellar structures, the ISM clouds have power-law size/mass distributions, and they exhibit power-law mass--size relations with fractional slopes \citep[e.g.][]{Elmegreen1996, RomanDuval2010}.

Supersonic turbulence has also been argued to play a dominant role in controlling the hierarchical substructures in the ISM \citep{Elmegreen2004}. In star-forming complexes, supersonic turbulence, aided by gravity, leads to the fragmentation of ISM into smaller and smaller substructures. \citet{Stanimirovic2001} analyzed the position--position--velocity data cubes of H~{\scriptsize I} lines in the SMC. They found that the observations agree with theoretical expectations of turbulence \citep{Lazarian2000}. Thus, the significant density fluctuations of H~{\scriptsize I} in the SMC \citep{Stanimirovic1999, Stanimirovic2000} are indeed caused by active turbulence instead of being a static structure. With power-spectrum analysis, \citet{Stanimirovic1999, Stanimirovic2000} found fractal dimensions of $D_2$~=~1.4--1.5 for H{\scriptsize~I} and dust in the SMC. These values are also consistent with that of the young stellar structures ($D_2$~=~1.48~$\pm$~0.03 from the mass--size relation, and $D_2$~=~1.4~$\pm$~0.1 from the size distribution; Sections~\ref{corr.sec} and \ref{dist.sec}).

Thus, the above discussion supports a scenario of hierarchical star formation in a turbulent ISM, which leaves imprints of the ISM properties on the young stellar structures. The remarkable similarities between the ISM and the stellar structures also suggest that the dynamical evolutionary effects are not significant. With dynamical evolution, the young stellar structures in the SMC, especially the unbound ones, will be dispersed on a timescale of $\sim$100~Myr (see the discussion in Section~\ref{ums.sec}).

The effect of turbulence is also consistent with the surface density distribution. Lognormal distributions have been observed for the column/volume densities of molecular clouds \citep{Lombardi2010, Schneider2012} or of simulational turbulent gas \citep{Klessen2000, Federrath2010, Konstandin2012}. With turbulence, such a distribution can be understood in a purely statistical way, if self-gravity and thermal pressure are unimportant (\citealt{VazquezSemadeni1994}; see also Section~4.4 of \citealt{Sun2017b}).  With star formation following gas distribution, the young stellar structures should also have a lognormal surface density distribution. This is in agreement with the result described in Section~\ref{dist.sec}.

Strong self-gravity, shocks, and rarefaction waves may lead to deviations from lognormal density distributions \citep{Klessen2000, Federrath2010}. Especially, strong gravity in star-forming clouds would cause a power-law tail at high surface densities \citep[see e.g.][]{Schneider2012}. However, our results do not show such deviations to any statistical significance. Such high-surface density structures (e.g. dense star clusters) usually have small sizes. And since our identification is complete only for scales larger than 10~pc, it is possible that the surface density distribution, shown as the histogram with error bars in Fig.~\ref{dist.fig} (bottom panel), does not reach the high surface densities of the power-law tail. While the all-structure distribution (histogram without error bars) does exhibit a number excess at high surface densities, we note again that this distribution is affected by incompleteness because structures with low surface densities and small sizes could be missing (see also Fig.~\ref{corr.fig}, bottom panel).

\subsection{Comparison with Other Works}
\label{comp.sec}

\citet{Gieles2008} investigated the spatial distribution of stars of various ages in the SMC. Using two-point correlation functions (TPCFs) and the $Q$-parameter \citep[introduced by][]{Cartwright2004}, they found that the SMC's overall stellar distribution evolves from a high degree of substructures ($\sim$10~Myr) to a smooth radial density profile. At $\sim$75~Myr the stellar distribution becomes statistically indistinguishable from a smooth distribution, suggesting that most of the young stellar structures have been erased by that age.

\citet{Bonatto2010} studied the size distributions of young (ages~$\lesssim$~10~Myr) clusters, old (ages~$\gtrsim$~600~Myr) clusters, and non-clusters (nebular complexes and their stellar associations) in the SMC. They were all found to follow power-law size distributions, with slopes of $-$3.6, $-$2.5, and $-$1.9, respectively. Note that their slopes are for the size distributions per {\it linear} interval, and correspond to $-$2.6, $-$1.5, and $-$0.9 for the distributions per {\it logarithmic} interval. These values differ from $\alpha(R)$~=~$-$1.4~$\pm$~0.1 derived in Section~\ref{dist.sec}. On the one hand, star clusters and associations are not distinguished and they are analyzed together in our work. On the other hand, their analysis is based on the catalog of extended sources of \citet{Bica2008}, which was compiled from extensive previous catalogs by different authors. As a result, we expect our analysis to be more systematic and less biased.

\citet{Gouliermis2014} reported a bimodal clustering of stars in the SMC star-forming region, NGC~346. They found that the TPCF is a double power law with a break at 20~pc. Based on simulations, they argued that the inner, flatter power law reflects a centrally concentrated stellar distribution (typical for star clusters), while the outer, steeper part corresponds to an extended, hierarchical component. Although their study focused on a smaller region in the SMC, they derived a fractal dimension for the hierarchical component of $D_2$~=~1.4 based on the slope of the outer TPCF (or a volume fractal dimension of $D_3$~=~2.4 based on their simulations). This value is consistent with what we have derived for the SMC globally.

\citet{Bastian2009} studied the young stellar structures in the LMC. They found that the structures have a power-law luminosity function (per linear interval) with a slope of 1.94~$\pm$~0.03. Assuming that the structures have no significant age spreads or stochastic sampling effects, this luminosity function corresponds to a structure mass distribution similar to that of the SMC young stellar structures (Section~\ref{dist.sec}). They also derived a fractal dimension of $D_2$~=~1.8 based on the $Q$-parameter. However, we note that this method is subject to large uncertainties \citep[see Fig.~5 of][]{Cartwright2004}.

In \citet{Sun2017a, Sun2017b} we studied young stellar structures in the LMC's 30~Dor and Bar complexes, respectively. Young stellar structures in the two complexes have similar properties to those in the SMC, in terms of the size/mass/surface density distributions and the mass--size relation. More quantitively, the fractal dimensions are $D_2$~=~1.6~$\pm$~0.3 for the 30~Dor complex and $D_2$~=~1.5~$\pm$~0.1 for the Bar complex. These values are also close to that for the SMC young stellar structures, irrespective of the different galactic environments for the Magellanic Clouds.

The hierarchical young stellar structures have also been investigated in Galactic star-forming regions and in some other galaxies \citep[e.g.][]{Larson1995, Simon1997, Elmegreen2001, Elmegreen2006, Elmegreen2014, Gouliermis2010, Gouliermis2015, Gouliermis2017}. The results are in general agreement with ours. Specifically, the typical fractal dimensions are close to 1.4--1.5, but \citet{Elmegreen2014} found that starburst dwarfs or H~{\scriptsize II} galaxies have larger projected fractal dimensions if they contain one or two dominant stellar complexes. We refer to \citet{Sun2017a} for an extensive comparison of the fractal dimensions.

\section{Summary and Conclusions}

In this paper we study the hierarchical young stellar structures in the SMC. This is based on the VMC survey and UMS stars selected from the ($J - K_s$, $K_s$) CMD. We apply a contour-based clustering analysis to the KDE map of the UMS stars. Young stellar structures are identified as overdensities on a series of significance levels. After rejecting less reliable detections, we identify 556 young stellar structures in total.

The young stellar structures are distributed in a hierarchical manner, such that larger structures on lower significance levels contain one or more smaller ones on higher significance levels. We illustrate this behavior with a dendrogram, which is a structure tree showing the ``parent--child" relations between the young stellar structures on different significance levels. The structures also have highly irregular morphologies. This is quantitively analyzed based on the perimeter--area relation of their (projected) boundaries, which suggests a perimeter--area dimension of $D_p$~=~1.44~$\pm$~0.02.

Size, stellar number (the number UMS stars within their boundaries, proportional to mass to a good approximation), and surface density are also analyzed statistically for the young stellar structures. There is a power-law correlation between stellar number and size with a slope of $\kappa$~=~1.48~$\pm$~0.03, while the surface density and size are not significantly correlated. The young stellar structures follow power-law size and mass distributions (per logarithmic interval) with slopes of $\alpha(R)$~=~$-$1.4~$\pm$~0.1 and $\alpha(N_\ast)$~=~$-$1.0~$\pm$~0.1, respectively. The surface density of the structures, on the other hand, exhibits a lognormal distribution.

These properties are remarkably similar to those of the ISM which is regulated by supersonic turbulence. Thus, our results support a scenario of hierarchical star formation, which leaves the imprints of the ISM properties on the young stellar structures. We also make a comparison with previous studies of young stellar structures.

\acknowledgements
We thank Jim Emerson for constructive suggestions on this paper. The analysis in this paper is based on observations collected at the European Organisation for Astronomical Research in the Southern Hemisphere under ESO program 179.B-2003. We thank the Cambridge Astronomy Survey Unit (CASU) and the Wide Field Astronomy Unit (WFAU) in Edinburgh for providing calibrated data products through the support of the Science and Technology Facility Council (STFC) in the UK. This work was supported by the National Key Research and Development Program of China through grant 2017YFA0402702. \text{N.-C.~S} and \text{R.~d.~G.} acknowledge funding support from the National Natural Science Foundation of China through grants 11373010, and U1631102. \text{M.-R.~L.~C.} and \text{C.~P.~M.~B.} acknowledges support from the European Research Council (ERC) under European Union's Horizon 2020 research and innovation programme (grant agreement No. 682115).

\end{document}